\documentclass[twocolumn]{aastex63}

\usepackage[T1]{fontenc}

\shorttitle{Vapor enrichment and Disk Structure}
\shortauthors{Kalyaan et al.}
\graphicspath{{./}{figures/}}

\begin{document}

\title{The Effect of Dust Evolution and Traps on Inner Disk Water Enrichment}

\correspondingauthor{Anusha Kalyaan}
\email{a\_k363@txstate.edu}

\author[0000-0002-5067-1641]{Anusha Kalyaan}
\affiliation{Department of Physics,
Texas State University, 749 N Comanche St,
San Marcos, TX 78666, USA}

\author[0000-0001-8764-1780]{Paola Pinilla}
\affiliation{Mullard Space Science Laboratory, University College London, Holmbury St Mary, Dorking, Surrey RH5 6NT, UK.}

\author{Sebastiaan Krijt}
\affiliation{School of Physics and Astronomy, University of Exeter, Stocker Road, Exeter EX4 4QL, UK}

\author{Andrea Banzatti}
\affiliation{Department of Physics, 
Texas State University, 749 N Comanche St,
San Marcos, TX 78666, USA}

\author{Giovanni Rosotti}
\affiliation{Dipartimento di Fisica, Università degli Studi di Milano, via Giovanni Celoria 16, 20133, Milano, Italy}

\author{Gijs D. Mulders}
\affiliation{Facultad de Ingenier\'{i}a y Ciencias, Universidad Adolfo Ib\'{a}\~{n}ez, Av. Diagonal las Torres 2640, Pe\~{n}alol\'{e}n, Santiago, Chile}
\affiliation{Millennium Institute for Astrophysics, Chile}

\author[0000-0001-9321-5198]{Michiel Lambrechts}
\affiliation{Center for Star and Planet Formation, GLOBE Institute, University of Copenhagen, Øster Voldgade 5-7, 1350 Copenhagen, Denmark }

\author[0000-0002-7607-719X]{Feng Long}
\altaffiliation{NASA Hubble Fellowship Program Sagan Fellow}
\affiliation{Lunar and Planetary Laboratory, University of Arizona, Tucson, AZ 85721, USA}

\author[0000-0002-7154-6065]{Gregory J. Herczeg}
\affiliation{Kavli Institute for Astronomy and Astrophysics, Peking University, Beijing 100871, China}
\affiliation{Department of Astronomy, Peking University, Beijing 100871, China}



\begin{abstract}

Substructures in protoplanetary disks can act as dust traps that shape the radial distribution of pebbles. By blocking the passage of pebbles, the presence of gaps in disks may have a profound effect on pebble delivery into the inner disk, crucial for the formation of inner planets via pebble accretion. This process can also affect the delivery of volatiles (such as H$_2$O) and their abundance within the water snow line region (within a few au). In this study, we aim to understand what effect the presence of gaps in the outer gas disk may have on water vapor enrichment in the inner disk. Building on previous work, we employ a volatile-inclusive multi-Myr disk evolution model that considers an evolving ice-bearing drifting dust population, sensitive to dust-traps, which loses its icy content to sublimation upon reaching the snow line. We find that vapor abundance in the inner disk is strongly affected by fragmentation velocity (v$_{\rm f}$) and turbulence, which control how intense vapor enrichment from pebble delivery is, if present, and how long it may last. Generally, for disks with low to moderate turbulence ($\alpha$ $\le$ 1 $\times$ 10$^{-3}$)
and for a range of v$_{\rm f}$, radial location, and gap depth (especially that of the innermost gaps), can significantly alter enrichment. Shallow inner gaps may continuously leak material from beyond it, despite the presence of additional deep outer gaps. We finally find that the for realistic v$_{\rm f}$ ($\le$ 10\,m\,s$^{-1}$), presence of gaps is more important than planetesimal formation beyond the snow line in regulating pebble and volatile delivery into the inner disk. 
\end{abstract}



\section{Introduction}\label{sec:intro}

Millimeter interferometric observations reveal that protoplanetary disks often exhibit substructure in dust emission in the form of gaps and rings \citep[e.g.,][]{andrews20,huang18a,long18}, and more recently, even in gas \citep[][]{teague18,kzhang21,wolfer22}. Gaps, potentially carved by the presence of planetary companions \citep[][]{paardekooper2007,rice2006,zhu12,marzari19,marzariangelo20}, play an important role in shaping the dust distribution and dynamics in disks by slowing the radially inward drift of dust particles and concentrating them in pressure bumps beyond gaps \citep[][]{pinilla2012} becoming observable ring-like structures in disks.

Dust trapping in rings beyond gaps, however, has consequences beyond the local disk region. By halting dust delivery, the presence of gaps in outer disk regions can create expansive solid reservoirs in the outer disk, leading to large observed dust disks with sizes spanning hundreds of astronomical units. In contrast, the rapid radial drift of dust in smooth unstructured disks likely make them compact, radially extending to only $\sim$ tens of astronomical units \citep[][]{huang18a,long18,long19,appelgren20,marelmulders21,toci21}. Moreover, the trapping of dust and pebbles in the outer disk can significantly impact planet formation and chemistry in the inner disk. Pebble mass-fluxes into the inner few astronomical units can dictate what types of planets can form via pebble accretion there, whether they may be Earths, super-Earths or even the cores of giant planets \citep[][]{lamb19}. Additionally, these incoming pebbles and dust particles carry significant masses of volatile ices (such as H$_2$O and CO) within them \citep[][]{pont14}. The  presence of any gaps and traps that halt their passage from the outer to the inner disk therefore can not only curtail the formation of larger planets in the terrestrial planet region, it can also reduce the overall mass of volatiles brought into the inner disk, where they may sublimate into vapor within the snow line region, affecting the local chemistry, the amount of volatiles that is accreted into forming planetesimals, and even the atmospheric compositions of forming planets \citep{cieslacuz06,morbi16,booth17, vent20,oberg21,schneiderb21,schneiderb21_2}. Inner disk volatile abundances can also provide a novel way to constrain pebble mass-fluxes into the terrestrial planet formation region \citep[][]{banzatti20,kalyaan21}.

To get a broader and more complete view of the delivery of ice-rich pebbles from the outer to the inner disk, it is critical to look beyond millimeter interferometry that informs us about the presence of gaps in the outer disk and radial drift there, and consider infrared spectra of molecules that inhabit the warm inner regions. Synthesizing insights from both types of observations, \citet{banzatti20} found an anti-correlation between disk size (and presence of substructure) and the water luminosity tracing the abundance of the water molecules in the inner disk. They found that smaller disks generally show higher H$_2$O luminosities (suggestive of higher H$_2$O column densities) than larger disks with substructure, indicating that pebble delivery regulated by the presence of gaps in the outer disk may be able to significantly affect the inner disk vapor abundance. This scenario was then modeled and verified with a volatile-inclusive disk evolution model in \citet{kalyaan21}, where the effect of a gap on inner disk vapor enrichment was explored. \citet[]{kalyaan21} modeled a radially-drifting ice-bearing single-sized dust population that lost ice to sublimation on approaching the water snow line in the inner disk. This work found that in general, inner disk water vapor abundance evolved with time. As pebbles rich in water ice drifted inwards into the inner disk, vapor mass increased, and peaked when the bulk of the icy pebble population was brought inward, and later declined with time as water vapor was accreted onto the star. They also found that the presence of a gap and its radial location  significantly influenced the inner disk vapor enrichment, with deep innermost gaps (outside of the snow line) drastically restricting the delivery of ice-rich pebbles into the inner disk.

In this work, we build on our previous modeling by incorporating a `full' evolving dust population (shaped by growth and fragmentation) via the two-population model by \citet{birnstiel12}, rather than the single-sized dust population previously assumed.  Previous studies have also used the two-population model or similar models to model the snow line region and volatile/ice transport and distribution in the disk. \citet[]{schoonormel17} used a characteristic particle method with an assumed particle size at each radius and time to explore whether a complex internal aggregate structure of drifting icy pebbles could foster planetesimal formation just inside and outside of the snow line. Similarly, \citet[]{drazali17} use the two-population model within a full disk evolution model to find that planetesimal formation is favored just beyond the snow line. Finally, \citet[]{booth17} and \citet[]{schneiderb21,schneiderb21_2} also used this model to explore the influence of pebble growth and drift on the chemical enrichment of inner disk gas and giant planet atmospheres. While the above studies collectively include detailed microphysics as well as planet formation and migration, to our knowledge, no study so far has systematically explored the effect of disk gaps, including their location and depth, on vapor enrichment in the inner disk with an evolving dust population model, as we present here. 

In this study, we perform detailed parameter studies on various disk and vapor/particle properties such as turbulent viscosity $\alpha$, diffusivities of vapor and dust, and fragmentation velocities of ice and silicate particles, gap properties such as their radial location and depth, and study the effect of each property on inner disk vapor enrichment. Moreover, we also consider new insights from very recent experiments performed at lower temperatures (more consistent with the outer solar nebula) that suggest that fragmentation velocities of icy silicates and pure silicates may be comparable to each other, i.e., $\sim$ 1\,m\,s$^{-1}$ \citep[]{gund18,musiowurm19}, rather than what was suggested from older experiments \citep{blumwurm2000,gundblum15,musio16} that icy particles were likely to be more stickier by at least an order of magnitude. We finally also study the effect of including planetesimal formation on vapor content in the inner disk. 

This paper is organized as follows. We describe the model used in this work in Section \ref{sec:methods} and delve into the results of our detailed parameter study in Section \ref{sec:results}. We discuss in detail the main insights gained from this work in Section \ref{sec:discussion} and present the main conclusions of our study in Section \ref{sec:conclusions}.

\section{Methods}\label{sec:methods}

\begin{figure*}[htb!]
	\centering
	\includegraphics[scale=0.55]{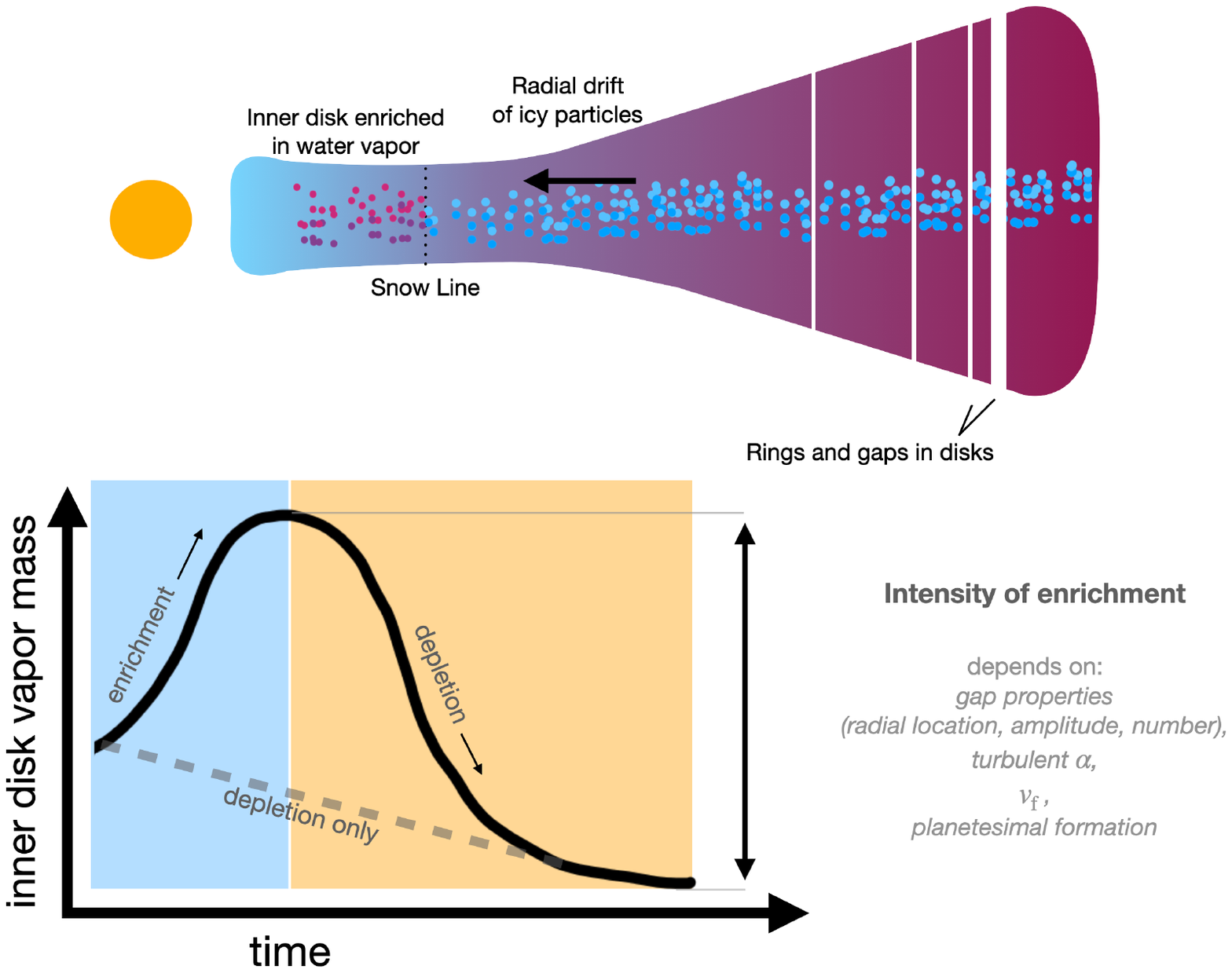}
	\caption{Schematic figure presenting our multi-Myr volatile-inclusive disk evolution model with disk structure (top).  Icy particles radially drift inwards, enriching the inner disk with water vapor. Vapor mass typically rises with incoming pebble drift, peaks and declines with time, as disk mass accretes onto the star (bottom).}
	\label{Fig:schematic}
\end{figure*}

The main motivation of this work is to model pebble dynamics in disks with gaps and assess the effect of structure on icy pebble delivery into the inner disk and the resulting vapor enrichment there. We use the multi-million year disk evolution model with volatile transport as described in \citet{kaldesch19}, supplemented by the addition of structure in the form of gaps as described in \citet{kalyaan21}. We then incorporate the two-population model from \citet{birnstiel12} to include a `full' dust population, with particle size at each location evolving through growth and fragmentation, and then finally also include planetesimal formation by using clumping criteria from recent streaming instability simulations (see Figure \ref{Fig:schematic}). In this section, we will describe the specific additions we have made here over the model described in \citet{kalyaan21}. We list all the model parameters used in Table \ref{table:1}. 

\begin{table*}\label{table:1}
\centering
\begin{tabular}{ |p{6cm}|p{2cm}|p{6cm}|  }
 \hline
 \bf{Model Parameter} & \bf{Symbol} & \bf{Parameter Value/Range}\\
 \hline
 
 \multicolumn{3}{|c|}{\bf{Disk Parameters}} \\
 \hline
 Inner disk radius   & R$_{\rm in}$    &    0.1 au\\
 Characteristic radius &   R$_{0}$  & 70 au\\
 Outer disk radius &  R$_{\rm out}$ & 500 au\\
 Initial disk mass    & M$_{\rm disk}$ & 0.05 M$_{\odot}$\\
 Initial gas surface density at R$_0$ = 70 au  & $\Sigma_0$ &  14.3 g cm$^{-3}$\\
 Initial gas surface density power law index &  & -1 \\ 
 Disk viscosity & $\alpha$ & 5 $\times$ 10$^{-4}$, \textbf{10$^{-3}$},  5 $\times$ 10$^{-3}$ \\
 Opacity (fine dust) & $\kappa$ & 5 cm$^2$ / g \\
 Simulation time &  & 5 Myr \\
 \hline
 \multicolumn{3}{|c|}{\bf{Dust Parameters}} \\
 \hline
 Initial monomer size & $a_{\rm p}$ & 0.1 $\mu$m\\
 Internal particle density (ices, silicates) & $\rho_{\rm s}$  & 1.5 g cm$^{-3}$\\
 Initial solid particle surface density & $\Sigma_{\rm icy dust}$, $\Sigma_{\rm dust}$ & 0.005 $\times$ $\Sigma_{\rm gas}$ g cm$^{-2}$\\
 Fragmentation velocity & v$_{\rm f}$ & 1, $\textbf{5}$, 10\,m\,s$^{-1}$ \\
 Schmidt number for diffusion & Sc$_{\rm p}$ & 0.3, \textbf{1.0}, 3.0\\
 \hline
 \multicolumn{3}{|c|}{\bf{Vapor Parameters}} \\
 \hline
 Initial concentration of water vapor & $c$ & 1 $\times$ 10$^{-4}$\\
 Schmidt number for diffusion & Sc$_{\rm v}$ & 0.3, \textbf{1.0}, 3.0\\
 \hline
 \multicolumn{3}{|c|}{\bf{Gap Parameters}} \\
 \hline
 Gap location & R$_{\rm gap}$ & 7, 15, 30, 60, 100 au\\
 Gap depth & $\alpha_{\rm gap}$ & 1.5 $\times$ $\alpha_{0}$ (shallower), \textbf{10 $\times$ $\alpha_{0}$}, 75 $\times$ $\alpha_{0}$ (deeper)($\alpha_{0}$= 1 $\times$ 10$^{-3}$)\\
 Gap width &  & 2 $\times$ scale height \\
 Time when gap instantaneously forms  & t$_{\rm gap}$ & 0 Myr\\
 \hline
\end{tabular}
\caption{Table of parameters used in our simulations. Bold values indicate fiducial model parameters.}
\end{table*}

\subsection{Gas transport}

We evolve a disk of mass 0.05 M$_\odot$ around a solar-mass star with the standard evolution equations from \citet{lbp74}, evolving surface density $\Sigma$ over a discretized model grid of 450 radial zones between 0.1 au to 500 au as follows: 

\begin{equation}
 \frac{\partial \Sigma} { \partial t} = \frac{1}{2 \pi r} \frac{\partial\, \dot{\rm M} }{\partial\, r} \,,
\end{equation}
where the rate of mass flow $\dot{\rm M}$ is given as:

\begin{equation}
  \dot{\rm M}=6 \pi r^{1/2} \frac{\partial}{\partial r} \left(r^{1/2}\Sigma \,\nu \right).
\end{equation}

The initial gas surface density $\Sigma(r)$ is equivalent to a power law with an exponential taper in the outer disk \citep[]{hartmann98} as:

\begin{equation}\label{initialsigmaeqn}
    \Sigma = \Sigma_0 \left( \frac{r}{r_0} \right)^{-1} \exp \left[ - \frac{r}{r_0} \right] \,\,  {\rm g}\, {\rm cm}^{-3}.
\end{equation}

Here, $\Sigma_0$ = 14.3 g cm$^{-3}$ at the characteristic radius r$_0$ = 70 au. Note that even though initial disk size is 500 au in our models, most of the disk mass is concentrated within r$_0$.
We assume turbulent viscosity $\nu$ of the disk follows the standard prescription $\nu$ = $\alpha$ $c_s^2/\Omega$, where $c_s$ is the local sound speed and $\Omega$ is the Keplerian angular frequency. We assume a typical $\alpha$ = 10$^{-3}$ throughout the disk (unless stated otherwise). As done previously in \citet[]{kaldesch19}, we also assume that the disk is heated both by stellar radiation and accretion heating, and denote the contribution of each as ${\rm T}_{\rm acc}$ for accretional heating, and ${\rm T}_{\rm pass}$ for passive starlight, combined as follows:

\begin{equation}
   {\rm T}(r) = \left[ {\rm T}_{\rm acc}(r)^4 + {\rm T}_{\rm pass}(r)^4 \right]^{1/4},
\end{equation}

where $r$ is the radial distance from the star. Here, ${\rm T}_{\rm acc}$ is given by: 

\begin{equation}\label{tempacceqn}
{\rm T}_{\rm acc}(r) = \left[ \frac{27}{128} \, \frac{k}{\mu \sigma} \, 
 \Sigma(r)^2 \, \kappa \, \alpha \, \Omega(r) \right]^{1/3},
\end{equation}

where $\Sigma(r)$, is the gas surface density in the disk, and $\Omega(r)$ the Keplerian angular frequency. $k$ is the Boltzmann constant, $\sigma$ is the Stephen-Boltzmann's constant, $\mu$ is the mean molecular weight of the bulk gas, while $\kappa$ is the fine dust opacity, assumed to be 5 cm$^2$/g. For accretion heating, the constant fiducial $\alpha$ is adopted throughout the disk (including within the gap, as we explain in detail below). 

As we did before in \citet[]{kalyaan21}, we assume ${\rm T}_{\rm pass}$ to be:

\begin{equation}\label{tempeqn2}
{\rm T_{\rm pass}}(r) = 118 \, \left( \frac{r}{ 1 \, {\rm au}} \right)^{-0.5} \, {\rm K}.
\end{equation}

We also do not allow the temperature to exceed 1400 K ($\approx$ silicate vaporization limit) in the innermost disk. We additionally impose a minimum temperature = 7 K in the outermost colder regions. 
We run all our simulations for 5 Myr. 

\subsubsection{Gap formation}
As we did previously in \citet{kalyaan21}, instead of directly modeling the formation of a planet carving a gap in the gaseous disk best done in 2D/3D, we incorporate instantaneous gap formation at $t=0$ (start of simulation) by implementing a Gaussian peak in the turbulent viscosity $\alpha$ profile as below:

\begin{equation}
    \alpha(r) = \alpha_{\rm 0} + (\alpha_{\rm gap} - \alpha_{\rm 0})\, \exp\,(-x^2) \,.
\end{equation} 
Here, $\alpha_{\rm 0}$ is the uniform turbulent viscosity $\alpha$ chosen throughout the disk, and $\alpha_{\rm gap}$ is the peak value of $\alpha$ imposed at the center of the gap, which creates a gap in the disk by depleting the gas surface density here. $x = (r - r_{\rm gap})$ / gap width, we assume gap width to be 2 $\times$ H, where $H$ is the gas scale height, given as H = c$_{\rm s}$/$\Omega$. Additionally, we test various `depths' of the gap by varying $\alpha$ gap as some factor $\times$ $\alpha_{0}$ ( = 1 $\times$ 10$^{-3}$). We choose three gap depths in our simulations: 1.5 $\times$ $\alpha_0$, 10.0 $\times$ $\alpha_0$ and 75.0 $\times$ $\alpha_0$. We find these depths to be equivalent to a range in depletion in $\Sigma_g$ in the gap of (i) a $\sim$ few times; (ii) one order; and (iii) two orders of magnitude, respectively. We anticipate that these ranges in gas depletion are also likely to be distinct enough that they may be observationally distinguishable (e.g., see CO depletions in high resolution MAPS images in Figures 18 and 20 in \citet{kzhang21} and Figure 4 in \citet{wolfer22}). These gap depths in turn likely correspond to the presence of planet of mass 33 M$_{\oplus}$, 0.3 M$_J$ and $\ge$ 1.0 M$_J$ respectively, within the gap, when compared to numerical simulations done by \citet{zhang18} (see their Figure 2).

Note that we use the Gaussian $\alpha$ peak only for creating and maintaining the gap in the gaseous disk, and otherwise assume that $\alpha$ is uniform for all other physical processes, including setting the thermal structure (Equation \ref{tempacceqn}), dust diffusion (Equation \ref{diffeqn}), as well as calculating the maximum particle size in the fragmentation regime (Equation \ref{eqn:afrag}).

We assume that streaming instability is the dominant mechanism of particle clumping leading to planetesimal formation \citep[]{baistone2010}. We also note that massive planets embedded in disks larger than $\sim$ 5 M$_J$ can interact with the gas, and may even lead to the formation of an eccentric disk, likely depending on the disk viscosity $\nu$ \citep[]{kleydirk06,ataiee13,teyss17,bae19}. Particles in such eccentric rings may undergo more fragmentation \citep[]{zsom11} and may not efficiently be trapped beyond the gap. We therefore make the assumption that the gap-forming planets in our simulations are much smaller than 5 M$_J$.

\subsection{Dust evolution and transport}\label{methods:dust}
To incorporate dust growth and fragmentation, we adapt the two-population model of \cite{birnstiel12}, which reproduces the results of detailed full numerical simulations \citep{birnstiel10} that include growth, fragmentation and radial drift of dust grains (with the advantage of being not as computationally expensive). Dust grains start from monomer sizes of 0.1 $\mu$m at the start of the simulation, and grow with time, their rate of growth dependent on the local dust density. They grow to a maximum size that is typically limited by either fragmentation in the interior of the disk or radial drift in the outer regions. The fragmentation limit is set by the fragmentation threshold velocity v$_{\rm f}$ that we assume to be independent of particle size, based on laboratory experiments \citep[][]{guttler10}. We assume that v$_{\rm f}$ is uniform throughout the disk, or varying at the snow line where particles lose their icy content, which allows us to account for a range of stickiness of icy silicates. We discuss this in more detail in the next subsection.

The general treatment of radial transport of dust particles (including advection-diffusion and radial drift) is similar to that implemented in \citet{kalyaan21} and \citet{birnstiel12}. We assume that diffusivities of the dust particles are similar to the gas diffusivity, fulfilled at the limit of St $<$ 1 \citep[see their Appendix A]{birnstiel12}. We also note that we use the same radially uniform $\alpha$ (1$\times$ 10$^{-3}$) for dust diffusion, as well as for calculating the maximum grain size in the fragmentation limit.

\subsection{Icy particles}\label{methods:icypart}
As done in \cite{kalyaan21}, we implement two particle populations - one entirely made of water ice, and the other made entirely of silicates, in order to simulate icy silicate particles that have an icy mantle around a silicate core. The ices in these particles sublimate when they drift inwards and eventually approach warmer temperatures surrounding the water snow line region in the disk (a few astronomical units from the star). In the previous model, we assumed that these two particle populations would be equally abundant and be identical in particle size, and therefore move similarly.  In the current model, we replicate the same effect by tracking the fraction of water ice in the total incoming particle mass at the snow line. We calculate $f_{\rm H_2O}$ as follows:

\begin{equation}
    f_{\rm H_2O} (r) = \frac{\Sigma_{\rm icy}}{\Sigma_{\rm icy} + \Sigma_{\rm dust}}.
\end{equation}

Then, we assume a fragmentation velocity threshold v$_{\rm f}$ based on our assumptions of relative strength of icy silicate particles compared with bare silicate particles. We allow for two possibilities: either that icy particles are stickier and have a higher v$_{\rm f}$ (up to an order of magnitude, i.e., 10\,m\,s$^{-1}$) compared to silicates (1\,m\,s$^{-1}$) or that they are both roughly comparable to each other. In the former case, we calculate f$_{\rm H_2O}$(r) to track the bulk water abundance in particles. If f$_{\rm H_2O}$(r) $<$ 0.1, then v$_{\rm f}$(r) is assumed to be 1\,m\,s$^{-1}$. If $f_{\rm H_2O}$(r) $>$ 0.1, then v$_{\rm f}$(r) is assumed to be 10\,m\,s$^{-1}$. In the latter case, we consider that v$_{\rm f}$ is uniform irrespective of distance from the star and consider a range of radially constant v$_{\rm f}$ ranging from 1\,m\,s$^{-1}$ - 10\,m\,s$^{-1}$. Overall, we still maintain a water ice abundance (by mass) in the incoming particle distribution that is roughly 50 $\%$ in ice and silicates in the outer disk beyond the snow line. (We note that our results do not change for any chosen $f_{\rm H_2O}$ threshold values $<$ 0.4, due to the steep decrease in $f_{\rm H_2O}$ at the snow line in our models.)

For both ice and silicates, we assume that $\rho_{\rm s}$ = 1.5 g cm$^{-3}$. 

\subsection{Planetesimal Formation}
We incorporate planetesimal formation as follows. We adopt the same prescription as in \citet{draz16} (their Equation 16) for computing the mass of planetesimals that form from dust at each radial location, but in place of their stricter conditions dictating where and when planetesimals can form, we adopt the latest criteria from streaming instability simulations by \citet{liyou21}. In this work, the authors find that the value of $Z$ (i.e., $\Sigma_{\rm d}/\Sigma_{\rm g}$) where strong clumping can take place, depends on the Stokes number St, and find that clumping can take place at lower Z values for a specific range of St, than previously thought. We use their prescribed fit from their Equation 11 (depicted in their Figure 4b) along with Equation 14. We use these criteria for planetesimal formation even in the pressure bump (which they are not intended for) due to a lack of similar criteria for pressure bumps \citep[e.g.][]{carrsimon22}, and admit that we may overestimate planetesimal formation at the pressure bump region. (In section \ref{results:planform}, we discuss in detail how planetesimal formation at the pressure bump region does not matter much for our results, as our traps block the passage of particles efficiently). 

We note that once planetesimals form, they are immobile and remain in place where they form, i.e., we do not consider any subsequent migration or dynamical evolution of planetesimals. They also do not accrete any pebbles. 

\section{Results}\label{sec:results}

In this work, we perform a detailed parameter study to investigate the effect of relevant disk/gap or vapor/particle parameters, such as gap location, gap depth, fragmentation velocity, viscosity $\alpha$, and particle and vapor diffusivity on the time-evolving vapor enrichment in the inner disk. We also explore the effect of including planetesimal formation, and the presence of multiple gaps. In each subsection, we go through the effect of systematically varying each of the above-mentioned parameters while keeping others constant (see Table \ref{table:1}), and discuss the most significant insights in Section \ref{sec:discussion}. (In all our simulations, we assume that the depth of the gap is given by 10 $\times$ $\alpha_0$ where fiducial $\alpha_0$ = 1 $\times$ 10$^{-3}$, unless stated otherwise. We also assume that the fiducial fragmentation velocity threshold v$_{\rm f}$ in simulations with dust growth and fragmentation is 5\,m\,s$^{-1}$.)

\subsection{Without Dust Growth and Fragmentation}

\begin{figure*}[htb!]
	\centering
	\includegraphics[scale=0.65]{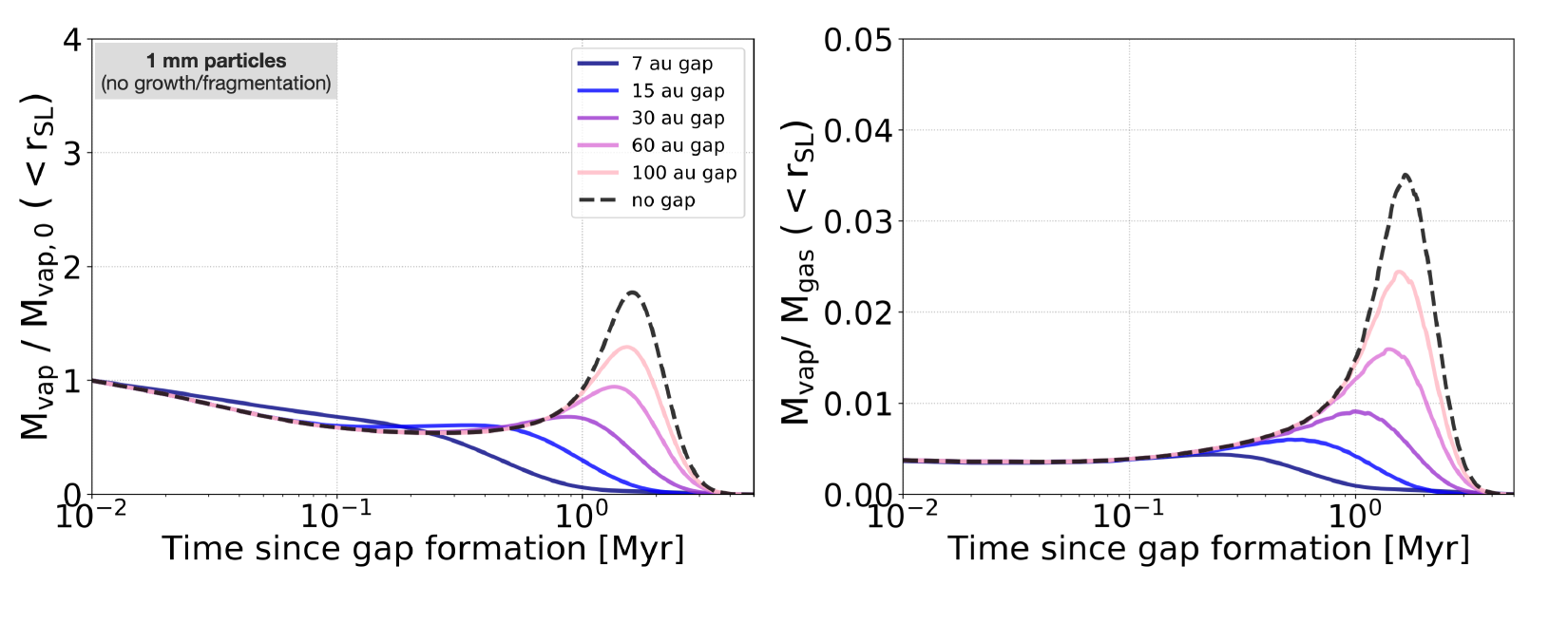}
	\caption{Left panel shows the time evolution of vapor enrichment in the inner disk, i.e., mass of water vapor within the snow line region at time t, normalized to mass of water vapor within the snow line at time t=0, for simulations without dust growth or fragmentation. Right panel shows time evolution of vapor abundance, i.e., mass of vapor within the snow line region normalized to mass of gas within the snow line region, for the same simulations. Different colors denote profiles for simulations where gap is located at different radii. Black dashed line shows the case with no gap for comparison. }
	\label{Fig:nogrowthfrag}
\end{figure*}

We begin our study by exploring the effect of including growth and fragmentation of dust particles in uniform disks (without gaps), and show the simulations without these physical processes in Figure \ref{Fig:nogrowthfrag} and with them in Figure \ref{Fig:gaplocation} (see black dashed lines). 

These results are plotted as `vapor enrichment' (i.e.,  mass of vapor present in the inner disk within the snow line region at any time, normalized to the mass of vapor present within the snow line at time $t=0$) in the left panel, and `vapor abundance' (i.e., mass of vapor with respect to the mass of bulk gas within the snow line at each time) in the right panel (see also Appendix \ref{app:vapgasrates}). Note that in our model, water vapor can only be present in the inner disk within the snow line. Moreover, since the temperature in the disk varies with time, the snow line also moves slightly inward with time (see appendix C and Figure 13 in \citet{kalyaan21}).

In simulations shown in Figure \ref{Fig:nogrowthfrag}, the entire dust population is composed of 1 mm particles. As discussed previously in  \citet{kalyaan21}, we find that the mass of vapor and its abundance in the inner disk evolves with time as icy pebbles drift inwards and deliver water vapor into the inner disk. Inner disk vapor mass and its abundance climbs and reaches a peak as bulk of the pebble mass enters the inner disk, and subsequently declines with time as stellar accretion takes over and depletes the inner disk of vapor.

In a uniform disk without gaps, the bulk of the same-sized dust population would drift inwards around the same time (here $\sim$ 2 Myr) depending on their radial location in the disk. On the contrary, including the processes of growth and fragmentation (see Figure \ref{Fig:gaplocation}) results in particles of a range of sizes. Larger pebbles drift more rapidly than smaller particles, leading to an earlier start in vapor enrichment, as well as a more intense but shorter enrichment episode. The intensity of enrichment may depend on how large the particles become before they fragment (see Section \ref{results:fragvel}).

In the following subsection, we will compare the effect of dust growth and fragmentation on vapor enrichment in disks with gaps.

\subsection{Varying Radial Location of Gaps}\label{sec:radiallocation}

\begin{figure*}[htb!]
	\centering
	\includegraphics[scale=0.65]{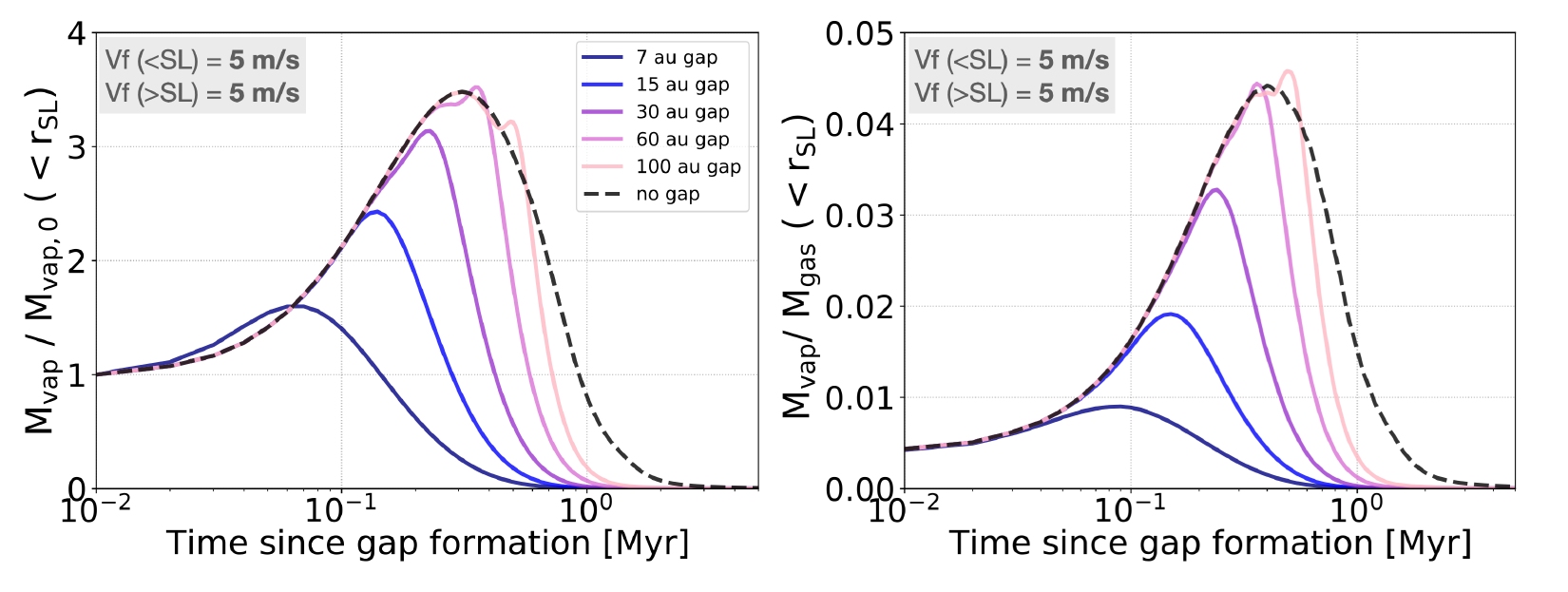}
	\caption{Left panel shows the time evolution of vapor enrichment in the inner disk, i.e., mass of water vapor within the snow line region at time t, normalized to mass of water vapor (within the snow line) at time t=0, for simulations with dust growth and fragmentation. Right panel shows time evolution of vapor abundance, i.e., mass of vapor within the snow line region normalized to mass of gas within the snow line region for the same simulations. Different colors denote profiles for simulations where gap is located at different radii. Black dashed line shows the case with no gap for comparison. Fiducial v$_{\rm f}$ = 5\,m\,s$^{-1}$ both inside and outside of the snow line is assumed. }
	\label{Fig:gaplocation}
\end{figure*}

We next vary the radial location of the gap in our simulations with and without the growth and fragmentation of dust particles.  As in the previous work \citep{kalyaan21}, we choose several gap locations from the inner to the outer disk, i.e., 7, 15, 30, 60 and 100 au from the star, to explore the fullest possible extent of the effect of structure on the mass of vapor in the inner disk. As we will explain below, we find that in spite of the initial disk size being 500 au, gaps farther than the critical radius have very little effect on inner disk vapor mass (in our case the gap at 100 au).

The results in both Figures \ref{Fig:nogrowthfrag} and \ref{Fig:gaplocation} follow the increase in mass of vapor brought to the inner disk, then peak of vapor enrichment when bulk of the drifting mass reaches the inner regions, followed by a depletion with time.  This profile is especially prominent in the case of a disk with no gap, where the peak vapor enrichment is highest from initial value. With a gap, however, both the peak value of vapor enrichment and the time it is attained may be different depending on where the gap is located. A closer-in gap (of equal depth) severely restricts the entry of ice-bearing pebble mass into the inner disk, as seen by the peak vapor enrichment for the disk with 7 au gap at only $\sim$ 1.6 at time $\sim$ 0.07 Myr in simulations with growth and fragmentation. At the same time, gaps farther out do not filter out as much pebble mass into the inner disk. This is because as they are further out, they do not block as much material as the gaps that are present closer-in. Moreover, pressure bumps farther out have a shallower pressure gradient, as the width of the gaps depends on the scale height at that radial location, which increases with $r$. Therefore, the further they are, the water enrichment profiles for these outer gaps more and more begin to resemble the no-gap scenario, both in value and time of peak vapor enrichment.

For a single-sized dust population, the radial location of the gap (for gaps of similar depths) determines the smallest particle size it is able to trap and block from passing through \citep[see Figure 12 in][]{kalyaan21}. With increasing radial distance $r$, the ability of a similar gap to block smaller particles gets progressively better. This is because smaller particles have larger St with increasing $r$, and are therefore easier to trap in pressure bumps. Additionally, they also diffuse less at greater distances (due to higher St). Therefore, for the same-sized particles, inner gaps are more leakier than outer gaps. This is consistent with earlier theoretical predictions from \citet{pinilla2012} that found a critical size of particles trapped in the pressure bump that decreased with increasing $r$, and is also consistent with observations that find a correlation between spectral index $\alpha_{\rm mm}$ and cavity radius \citep[]{pin14}. 

When the physics of particle growth and fragmentation is included, even if the gap is present in the drift-dominated regime (applicable for our outer gaps between ~15-100 au for fiducial v$_{\rm f}$), the fragmentation limit is always (eventually) reached in the pressure bump, replenishing the population of small particles within the bump. As outer gaps are better at trapping smaller particles, more of the small particles pass through the inner gap than through the outer gaps. The inner gaps are therefore slightly leaky compared to the outer gaps. (See also Appendix \ref{app:leakygaps} and Figure \ref{Fig:leakygap_app}).

Overall, for the same gap depth, vapor enrichment in the inner disk is affected by where the gap is located in the outer disk. If it is present too far out, it may not block enough icy particles to have any effect. An inner gap (in spite of being slightly leaky) may still block the most amount of icy particles outside of the gap. 

We also note the presence of very small surges in water vapor enrichment that slightly overshoot the no-gap profile, most noticeable for gaps in the outer disk ($>$ 60 au). These small surges arise from keeping the initial disk mass the same across all simulations whether or not gaps are present. Therefore, in the case of the disk with a gap further out, disk mass inside of the gap is higher in comparison to the no-gap case \footnote{We also find these small surges to be present in at least one other study \citep[]{andama22}}. In this work, we keep the magnitude of these surges negligible by keeping the depth of the gaps in most of the simulations in this work limited to 10 $\times$ $\alpha_0$. As we will explain later in Section \ref{results:gapstrength}, choosing a higher gap depth has little to no impact on the resulting time-evolving vapor enrichment in the inner disk. From here, we only show the vapor enrichment plots throughout the rest of the paper. This is because the snow line region moves inward with time, making the mass of gas within the snow line a constantly varying quantity. We therefore use `vapor enrichment', i.e., mass of vapor normalized to its value at initial time, rather than vapor-gas abundance, for the purpose of comparing across simulations (see Appendix \ref{app:vapgasrates}).

\subsection{Varying Gap Depth}\label{results:gapstrength}

\begin{figure*}[htb!]
	\centering
	\includegraphics[scale=0.62]{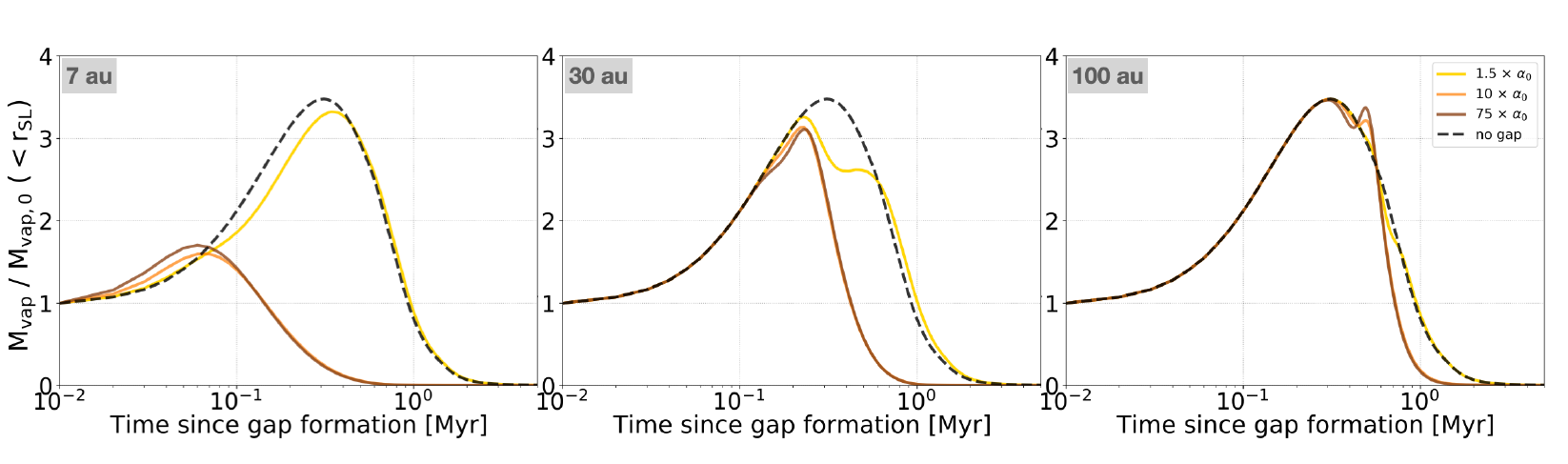}
	\caption{Time evolution of vapor enrichment for different gap depths at three different radial locations: 7 au (left panel), 30 au (middle panel) and 100 au (right panel). Different colors denote different gap depths (yellow denotes a shallower gap, brown denotes a deeper gap than fiducial). Black dashed line shows the case with no gap for comparison. Fiducial v$_{\rm f}$ = 5\,m\,s$^{-1}$ both inside and outside of the snow line is assumed.}
	\label{Fig:gapstrength}
\end{figure*}

For specific radial gap locations, we also vary the depth of the gap by parameterizing the turbulent $\alpha$ at the center of the gap as a factor $\times$ $\alpha_0$. As mentioned before, we select three gap depths 
: 1.5 $\times$ $\alpha_0$, 10 $\times$ $\alpha_0$ 75 $\times$ $\alpha_0$ for gaps located at 7, 30 and 100 au, as seen in Figure \ref{Fig:gapstrength}. We find that varying gap depth matters most for the innermost gaps, as seen in the case of 7 au (left panel). At this gap location, a shallow gap (1.5 $\times$ $\alpha_0$) makes for a highly inefficient barrier that allows for passage of material for several Myr after gap formation. For any values of gap depth $\ge$ 10 $\times$ $\alpha_0$, the gap becomes very efficient at blocking the passage of dust material and the water enrichment profiles become identical. Other gap locations (e.g., 30 au or 100 au shown in the middle and right panels) already show little deviation from the no-gap water enrichment profiles simply due to the smaller amount of ice-rich material they block based on their location. Over and above this, a shallower gap yields a very small deviation in the vapor enrichment profile relative to that seen with deeper gaps in the outer disk. In the case of shallow gaps, small particles pass through them (see Appendix \ref{appsec:gapstrength}). This leakage of material results in the high vapor enrichment peaks for simulations with shallow gaps. 

Overall, varying gap depth has the strongest effect on the innermost gaps. A deep inner gap can block a lot of ice-rich dust material from entering the inner disk, while a very shallow gap can let small particles pass through, yielding enrichment profiles that resemble that of a disk without any gaps.

\subsection{Varying Fragmentation Velocity}\label{results:fragvel}

\begin{figure*}[htb!]
	\centering
	\includegraphics[scale=0.65]{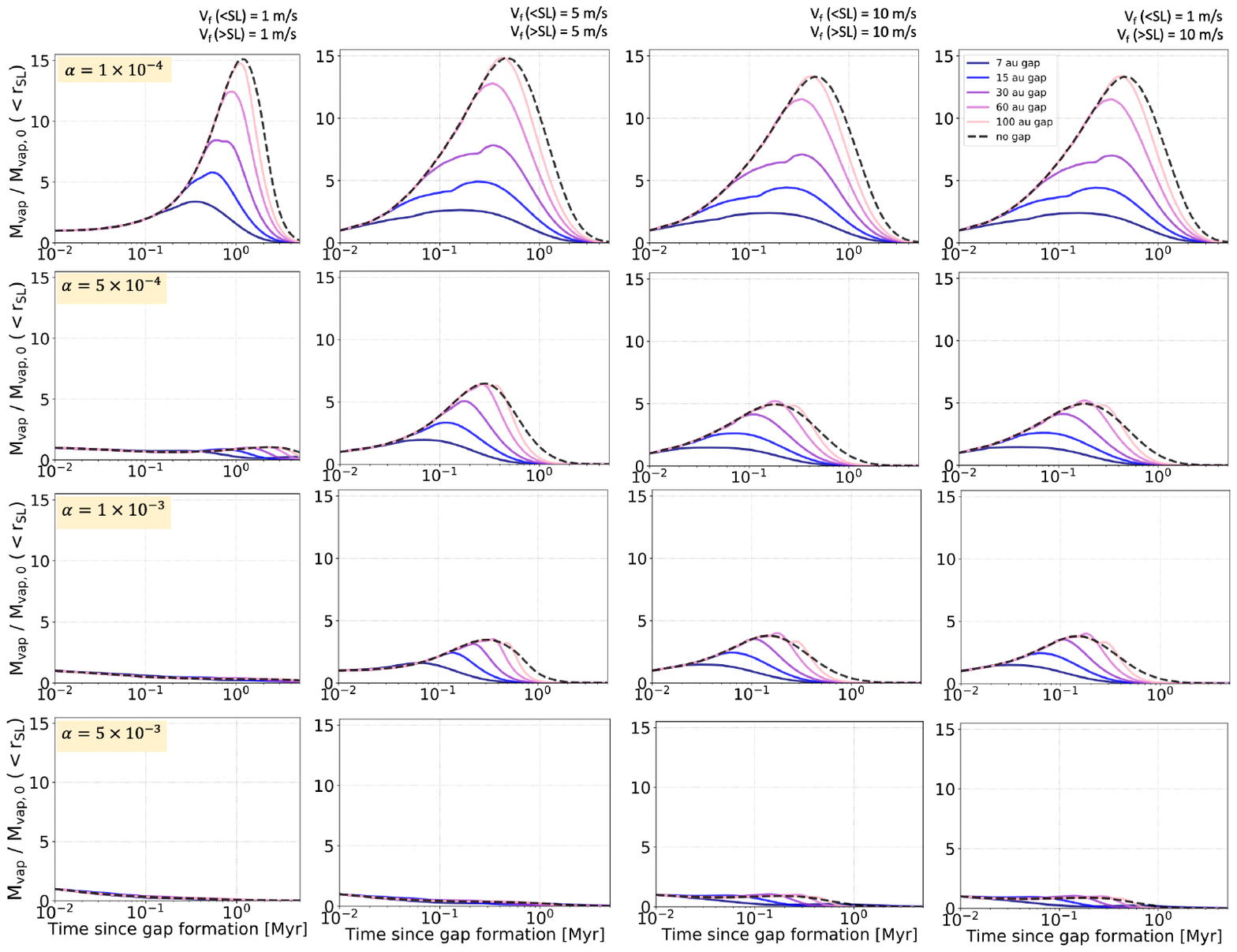}
	\caption{Grid of simulations performed for a range of fragmentation velocity v$_{\rm f}$, and turbulent $\alpha$. Rows from top to bottom show simulations for $\alpha$ = 1 $\times$ 10$^{-4}$, 5 $\times$ 10$^{-4}$, 1 $\times$ 10$^{-3}$ and 5 $\times$ 10$^{-3}$, respectively. First three columns from left to right show v$_{\rm f}$ = 1\,m\,s$^{-1}$, 5m/s and 10\,m\,s$^{-1}$ (for both inside and outside the snow line). Last column shows case with v$_{\rm f}$ = 1m/s inside the snow line, and 10\,m\,s$^{-1}$ outside of it. Colors and lines are as in Figure \ref{Fig:gaplocation}.  Note that Figure \ref{Fig:gaplocation}a is reproduced in the second panel of the third row to show the complete grid.}
	\label{Fig:vf_alpha}
\end{figure*}
As mentioned before, we take into account a range of assumptions for the relative fragmentation velocity thresholds v$_{\rm f}$ for ices and silicates in this work. We consider two different cases: i) that icy particles are stickier than silicate particles, and therefore have v$_{\rm f}$ = 10\,m\,s$^{-1}$ (an order of magnitude higher than silicate particles, which have v$_{\rm f}$ = 1\,m\,s$^{-1}$); and ii) that icy and silicate particles both have comparable tensile strengths against collisions, and therefore have similar v$_{\rm f}$. In this later case, as explained before, v$_{\rm f}$ is assumed identical for ices and silicates, i.e., constant both inside and outside of the snow line. Here, we test a range of constant v$_{\rm f}$ values: 1, 5 and 10\,m\,s$^{-1}$. We show these results in the third row of Figure \ref{Fig:vf_alpha}, for the fiducial value of turbulent $\alpha$ = 1 $\times$ 10$^{-3}$, where the first three columns correspond to v$_{\rm f}$ = 1, 5\,m\,s$^{-1}$ and 10\,m\,s$^{-1}$ for both ices and silicates, respectively, and the last column corresponds to the case where v$_{\rm f}$ = 10\,m\,s$^{-1}$ for ices, and v$_{\rm f}$ = 1\,m\,s$^{-1}$ for silicates. As before, for each case, we perform simulations with gaps at the same radial locations. 

We find that for a constant v$_{\rm f}$(r) = 1\,m\,s$^{-1}$, water vapor enrichment profiles show negligible deviation between simulations with or without a gap, irrespective of gap location, and simply decrease with time over 5 Myr from their initial value at t=0. For a constant v$_{\rm f}$(r) $\ge$ 5\,m\,s$^{-1}$, vapor enrichment profiles show the familiar increase, peak and decrease as drifting pebbles deliver ice-rich material into the inner disk, and the ice sublimates to vapor at the snow line. With a gap, as we found before, the vapor enrichment from initial time can be significantly lower or slightly lower relative to the no-gap scenario, depending on gap location.

Vapor enrichment profiles from simulations with radially constant v$_{\rm f}$ values of 5\,m\,s$^{-1}$ and 10\,m\,s$^{-1}$ vary in the following two ways. Higher v$_{\rm f}$ can result in a slightly higher no-gap peak i.e., 4 $\times$ initial value for 10\,m\,s$^{-1}$, compared to 3.5 $\times$ initial value for 5\,m\,s$^{-1}$). Higher v$_{\rm f}$ results in earlier peak times: ranging from $\sim$ 0.1 Myr after gap formation for the no-gap case to 0.02 Myr for 7 au gap for 10\,m\,s$^{-1}$, compared to 0.3 Myr for no-gap to 0.06 Myr for 7 au gap for 5\,m\,s$^{-1}$, with the peaks for the cases with other gap locations falling in between these two times for each v$_{\rm f}$. These variations in vapor enrichment for different v$_{\rm f}$ values occur because the maximum particle size at the fragmentation limit a$_{\rm frag}$ is given as follows \citep[equation 8]{birnstiel12}: 

\begin{equation}\label{eqn:afrag}
    a_{\rm frag} \propto \frac{2}{3 \pi} \frac{\Sigma}{\rho_{\rm s} \alpha} \frac{v^2_{\rm f}}{c^2_{\rm s}},
\end{equation}

where $\Sigma$ is the gas surface density, $\rho_{\rm s}$ is the internal particle density, and c$_s$ is the sound speed. Alternatively, the Stokes number at the fragmentation limit St$_{\rm f}$ is given by:

\begin{equation}\label{eqn:stfrag}
    {\rm St}_{\rm f} = \frac{1}{3} \frac{v^2_{\rm f}}{\alpha \,c^2_{\rm s}},
\end{equation}

Here, a$_{\rm frag}$ and ${\rm St}_{\rm f}$ are proportional to v$^2_{\rm f}$. As v$_{\rm f}$ is increased, small particles are able to grow to larger and larger sizes before they fragment. These larger particles are able to drift inwards more rapidly and also bring in more water with them more quickly as they drift, therefore showing higher and earlier peak vapor enrichment with increasing v$_{\rm f}$. (In all our simulations, the inner few astronomical units are always in the fragmentation-dominated regime.)

The case with 1\,m\,s$^{-1}$ is exceptional as such a low v$_{\rm f}$ allows for very little growth of particles. Most of the disk is at the fragmentation-dominated limit. These particles drift very slowly (slower than gas accretion into the star) that the inner disk is never really enriched with water vapor over the value at t=0, causing the vapor enrichment to decline continuously with time (see also Appendix \ref{app:vapgasrates}).
Finally, we explore different v$_{\rm f}$ for ices and silicates, and find these results to be identical to that of constant v$_{\rm f}$ = 10\,m\,s$^{-1}$. This is because the mass of water that is brought inwards in both of these cases is the same as what matters most in our simulations is the fragmentation velocity of icy particles.  

Overall, for moderate values of turbulent $\alpha$, lower v$_{\rm f}$ ( = 1\,m\,s$^{-1}$) yields similar vapor enrichment profiles, whether or not gaps are present. Higher v$_{\rm f}$ show distinct differences in these profiles that are dependent on the presence of a gap and their radial location. 

\subsection{Varying Viscosity}

Figure \ref{Fig:vf_alpha} shows not only the full simulation grid with various gap locations we performed for a range of v$_{\rm f}$ values but it also includes simulations performed for a range of turbulent viscosity $\alpha$. Overall, we explored four values of $\alpha$ ranging over an order of magnitude (1 $\times$ 10$^{-4}$, 5 $\times$ 10$^{-4}$, 1 $\times$ 10$^{-3}$ and 5 $\times$ 10$^{-3}$), suggested from observations \citep[see Table 3 in][]{rosotti23}, to study how low or high $\alpha$ may impact vapor enrichment in the inner disk. Results for these additional values of $\alpha$ are shown in the first, second and fourth rows of this grid. 

For lower $\alpha$ (than fiducial), we find that vapor enrichment peaks are higher than with fiducial $\alpha$, reaching up to 5 - 6.5 for $\alpha$ of 5 $\times$ 10$^{-4}$, and reaching up to 14 - 15 for $\alpha$ of 1 $\times$ 10$^{-4}$  for  the no-gap profiles with v$_{\rm f}$ = 5 and 10\,m\,s$^{-1}$. Even the 1m/s simulations show an increase in vapor enrichment over initial time. While the enrichment episode is only slight for $\alpha$ = 5 $\times$ 10$^{-4}$, it is substantially higher peaked for lower $\alpha$ = 1 $\times$ 10$^{-4}$, before decreasing eventually.

In contrast, for higher $\alpha$ (= 5 $\times$ 10$^{-3}$), vapor enrichment declines from initial time for all v$_{\rm f}$ values. Only the cases with v$_{\rm f}$ = 10\,m\,s$^{-1}$ show that vapor enrichment in the disk stays at $\sim$ 1 for 0.2-0.3 Myr after gap formation, and then decreases with time between 0.1 - 1 Myr depending on if there is a gap and where it is located.   

These effects can again be attributed to how $\alpha$ is inversely proportional to a$_{\rm frag}$ and St$_{\rm frag}$ in Equations \ref{eqn:afrag} and \ref{eqn:stfrag}. Higher $\alpha$ therefore results in a lower fragmentation size limit, and slower drift for particles in the disk, and vice versa.

\subsection{Varying Particle/Vapor Diffusivity}
\begin{figure*}[htb!]
	\centering
	\includegraphics[scale=0.5]{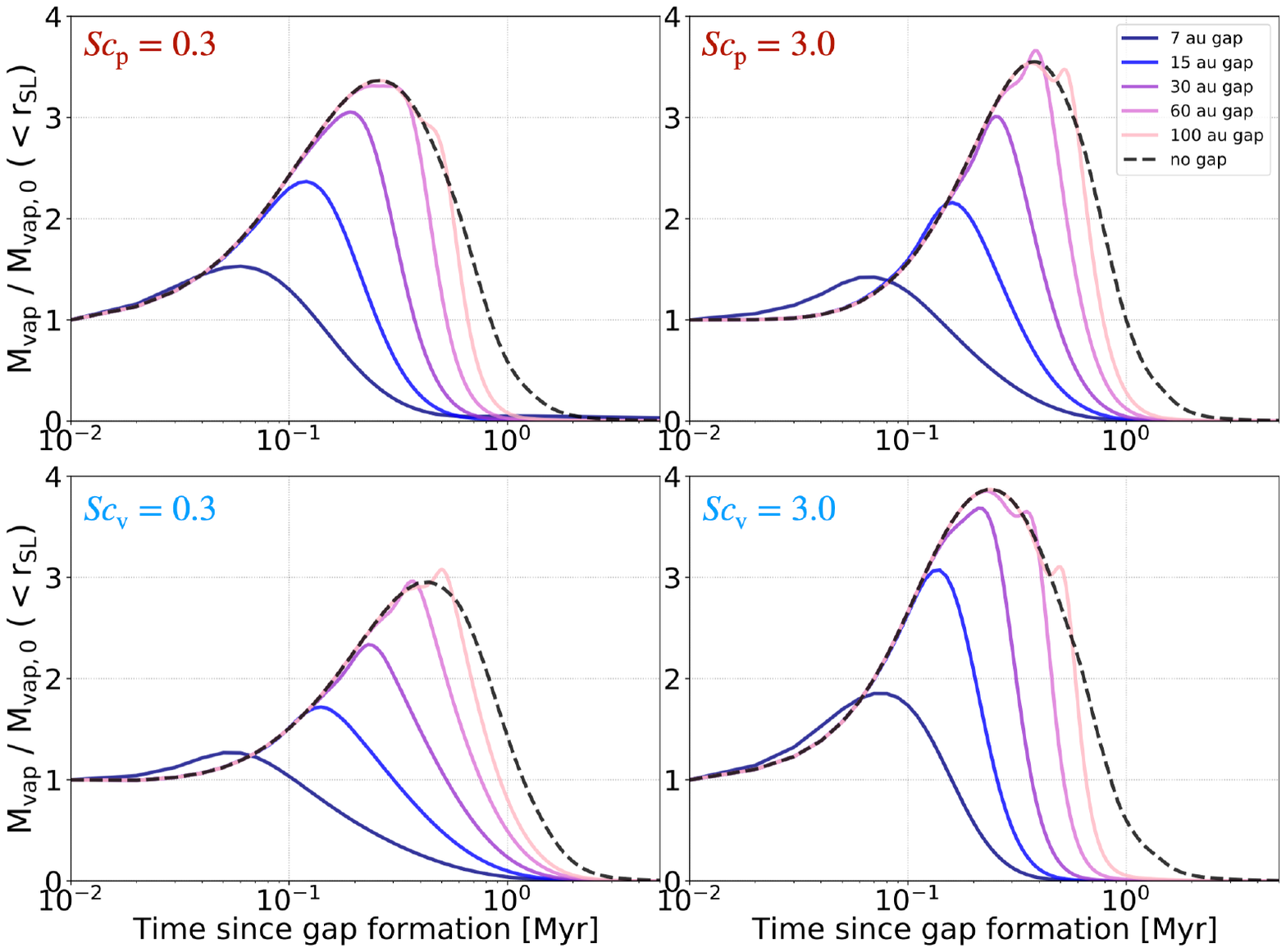}
	\caption{Time evolution of vapor enrichment for simulations performed with a range of particle and vapor diffusivities. Top row shows simulations varying particle diffusivity; bottom row shows simulations varying vapor diffusivity. Fiducial v$_{\rm f}$ = 5\,m\,s$^{-1}$ both inside and outside of the snow line is assumed. Colors and lines are as in Figure \ref{Fig:gaplocation}.}
	\label{Fig:partvapdiff}
\end{figure*}

We also perform some simulations where we study the effect of varying the diffusivities of particles and vapor, while keeping the fiducial value of $\alpha$. For particles, diffusivity $\cal{D}_{\rm part}$ is generally taken to be $\cal{D}_{\rm part}$ = $\cal{D}_{\rm gas}$/(1 + St$^2$). In our simulations, St is always $<<$ 1, and reaches a maximum of 0.1 in the outer disk at few tens of au. We therefore take $\cal{D}_{\rm part}$ $\approx$ $\cal{D}_{\rm t}$, which is given as:

\begin{equation}\label{diffeqn}
    \cal{D}_{\rm t} = \frac{\nu}{{\rm Sc}_{\rm t}}.
\end{equation}
Here, $\nu$ is the viscosity of the bulk disk gas, and Sc denotes the Schmidt number of the tracer $t$ in the bulk gas, which refers to either particles or vapor in the disk. We vary Sc for particles as well as for vapor (i.e., Sc$_{\rm p}$ and Sc$_{\rm v}$ respectively) over an order of magnitude around the fiducial value of 1 and show the results of our simulations in Figure \ref{Fig:partvapdiff}, where the top row shows simulations varying Sc$_{\rm p}$ and the bottom row shows simulations varying Sc$_{\rm v}$. Here, varying Sc$_{\rm p}$ (or equivalently $\cal{D}_{\rm part}$) physically implies that particles are more or less diffusive in the gaseous medium as they drift inward in the inner disk. Upon reaching the inner disk within the snow line, the ice in these particles sublimate to form vapor that yield these vapor enrichment profiles. Varying Sc$_{\rm v}$, on the other hand, takes into account how diffusive vapor is in the bulk gas after it is generated at the snow line and moves through the inner disk within the snow line region. As lower Sc implies higher diffusivity, the panels on the left (both Sc$_{\rm p}$ and Sc$_{\rm v}$ simulations) show that the vapor enrichment profiles are more shallow and spread out in time, compared to the profiles in the panels on the right, which show generally higher peak enrichments for the no-gap and gap simulations. However, varying Sc$_{\rm v}$ seems to have a more significant impact (with peak enrichments reaching $\sim$ 4.0 for Sc$_{\rm v}$ = 3.0 over $\sim$ 3.0 for Sc$_{\rm v}$ = 0.3) than varying Sc$_{\rm p}$ which shows comparatively little change (with peak enrichments reaching $\sim$ 3.5 for Sc$_{\rm p}$ = 3.0 compared to $\sim$ 3.3 for Sc$_{\rm p}$ = 0.3) in spite of a change of an order of magnitude in  Sc$_{\rm p}$.

Overall, the effect of changing vapor or particle diffusivity is relatively small; particle size set by $\alpha$ through fragmentation is more important in determining whether dust is trapped. 

\subsection{Including Planetesimal Formation}\label{results:planform}

\begin{figure*}[htb!]
	\centering
	\includegraphics[scale=0.65]{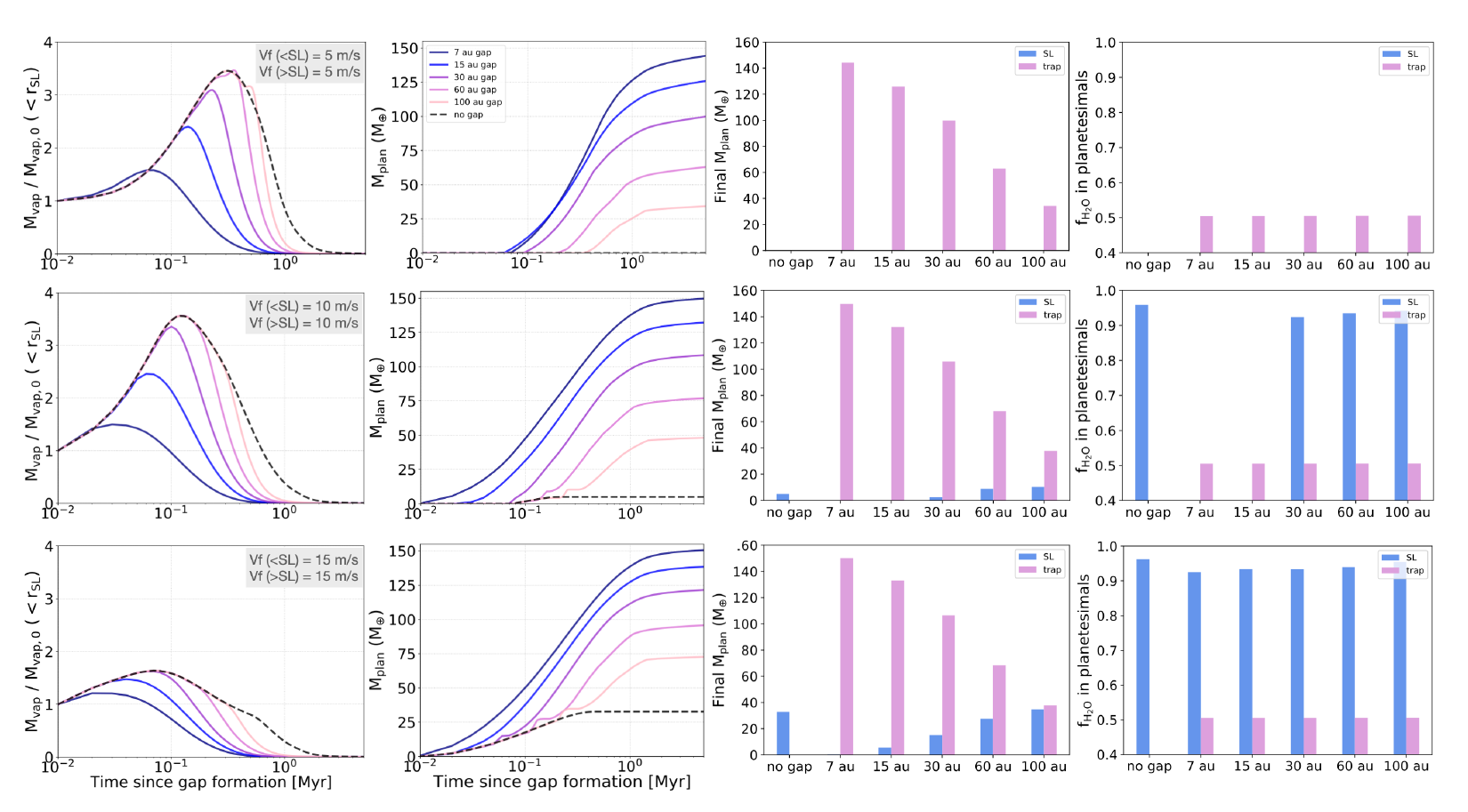}
	\caption{Plots with results of simulations with planetesimal formation for fiducial initial disk mass M$_{\rm disk}$ = 0.05 M$_{\odot}$. First column shows time evolution of vapor enrichment for same simulations performed with planetesimal formation. Second column shows corresponding total mass of planetesimals formed over time. Third column shows final planetesimal mass formed at the trap beyond the gap or the snow line (SL) region, for each simulation with gap at different radial locations. Fourth column shows the final fraction of water ice in planetesimals formed at the snow line or the dust trap. Top, middle and bottom rows correspond to v$_{\rm f}$ = 5 , 10 and 15\,m\,s$^{-1}$, inside and outside of the snow line. Line colors are as in Figure \ref{Fig:gaplocation}.}
	\label{Fig:planform1}
\end{figure*}

To study the effect of including the physics of planetesimal formation in our simulations on the vapor enrichment in the inner disk, we perform a series of simulations with a range of v$_{\rm f}$, where we assume that v$_{\rm f}$ both inside and outside of the snow line is 5, 10 or 15\,m\,s$^{-1}$ (shown in top, middle and bottom rows of Figure \ref{Fig:planform1}, respectively). The choice of v$_{\rm f}$ here plays an important role as it sets how big particles can grow before they fragment. The size of the particles (or equivalently the Stokes number of the particles) is critical for fulfilling the criteria for strong clumping as explored in \citet[]{liyou21}. As v$_{\rm f}$ is increased, more planetesimal formation takes place in regions where adequate dust-to-gas ratios of sufficiently large particles are reached. In our simulations, these conditions are satisfied in either one or two of the following regions: (i) in the pressure bump beyond the gap where particles drifting inwards from the outer disk are continuously being accumulated, grown and trapped; and (ii) just beyond the snow line where there is an overdensity of ice mass in particles due to retro-diffusion of vapor out through the snow line \citep[][]{rosjohan13,schoonormel17}.   
We show vapor enrichment profiles on the first column of Figure \ref{Fig:planform1}, corresponding to total mass of planetesimals formed (in the entire disk) over time for each simulation in the second column, the total planetesimal mass formed at 5 Myr at either the snow line region or pressure bump beyond the gap in the third column, and finally the fraction of water ice in planetesimals formed at either location in the fourth column. We also additionally show when planetesimal formation takes place beyond the snow line or at the bump in Appendix \ref{app:planform} and Figure \ref{Fig:planform_app3}.

For v$_{\rm f}$ = 5\,m\,s$^{-1}$,  we find that the vapor enrichment profiles (top-left subplot) are identical to the corresponding simulations without planetesimal formation shown earlier in Figure \ref{Fig:gaplocation}. In these simulations, conditions for strong clumping and planetesimal formation are only reached at the pressure bump beyond the gap. These particles are otherwise trapped beyond the gap anyway throughout the simulation. Therefore, there is no effect on pebble delivery as well as on vapor enrichment in the inner disk. For these simulations, there are no planetesimals formed in the disk for the no-gap simulation. But for disks with a gap, the final planetesimal mass formed beyond each gap decreases with increasing radial distance of the gap. This is because more dust mass can accumulate and grow beyond the gap if the gaps are closer-in. The total final planetesimal mass beyond a disk with a 7 au gap is $\sim$ 150 M$_{\oplus}$, decreasing to $\sim$ 40 M$_{\oplus}$ for a disk with a gap at 100 au. Furthermore, planetesimals form in the trap between 0.1 - 1 Myr for all these simulations. 

For simulations with v$_{\rm f}$ = 10\,m\,s$^{-1}$, a small mass of planetesimals ($<$ 10 M$_{\oplus}$)form after 0.07 Myr just beyond the snow line even for the case with no gap. 
Among the gap simulations, we find that for simulations with an inner gap at 7 or 15 au, planetesimals form only beyond the pressure bump. For simulations with an outer gap (30, 60 or 100 au), we find that planetesimals form beyond the pressure bump as well as just beyond the snow line region. In these three cases, the peak enrichments ($\sim$ 3.2 to 3.5) are smaller than that of corresponding simulations without planetesimal formation ($\sim$ 3.5 - 4.0), as some icy material is sequestered in planetesimals just beyond the snow line that would have otherwise been delivered into the inner disk and enriched it with water vapor. It is also important to note that a farther out gap allows for some planetesimal formation at the snow line (up to $\sim$ 10 M$_{\oplus}$ for the 100 au gap), rather than a close-in one. In these simulations, planetesimals begin to form at different times for disks with gaps at different locations (see Appendix \ref{app:planform} Figure \ref{Fig:planform_app3}). Planetesimal formation at the pressure bump can begin as soon as the gap is formed (i.e., t = 0) for a closer-in gap (7 au), and as late as 0.4 Myr for an outer gap (100 au). Planetesimal formation at the snow line only proceeds for a very short duration ($\sim$ 10,000 yr) at around 0.1 Myr. For the dust trap, on the other hand, it can begin at the start of the simulation (t=0), for the disk with a closer-in 7 au gap and proceed until 1 Myr after gap formation, or start at around 0.35 Myr and proceed until 1 Myr after gap formation, for a disk with an outer gap at 100 au.

Finally, we additionally explore an even higher value of v$_{\rm f}$ = 15\,m\,s$^{-1}$ and find that for all simulations with and without a gap, mass of planetesimals formed at the dust trap is similar to the previous case (v$_{\rm f}$ = 10\,m\,s$^{-1}$). However, additionally, a significant mass of planetesimals forms just beyond the snow line, as well just as beyond the gap, if a gap is present. The amount of mass locked up in planetesimals just beyond the snow line is sufficient to cause a drastic decrease in the vapor enrichment profiles in the inner disk for all runs with different gap locations, such that they do not peak beyond $\sim$ 1.6 for even the no-gap and outer gap simulations.

Generally, we find that similar masses of planetesimals are formed in dust traps for a range of v$_{\rm f}$, and that higher v$_{\rm f}$ allows more planetesimal formation at the snow line. We also find that planetesimals formed at the pressure bump beyond the gap have a 50-50 \% ice-to-rock ratio, in comparison to ice-rich planetesimals that form just beyond the snow line with 90 \% water ice via retro-diffusion or the cold-finger effect. This is consistent with results obtained from previous studies \citep[]{stevlun88,cuzzizahn04,rosjohan13} as well as recent work exploring the origin of CO-ice-rich comets beyond the CO snow line \citep[]{mousis21}. Other studies have also investigated planetesimal formation at the snow line. In a smaller disk, \citet{schoonen18} found icy planetesimals that formed beyond the snow line dominated the total mass ($\sim$ 100 M$_{\oplus}$) when compared with rocky planetesimals ($\sim$ 1 M$_{\oplus}$) that formed within the snow line region. They also found that the planetesimal formation proceeded for around 1000 - 10000 years. \citet{lich21} found two distinct reservoirs of planetesimals that could form by the outward and subsequent inward migration of the snow line across the disk. They argued that one reservoir forms mainly by the cold-finger effect at 1.3-7.5 au from 0.2 - 0.35 Myr and is composed of $\sim$ 1 M$_{\oplus}$, and a second reservoir of planetesimals of $\sim$ 300 M$_{\oplus}$ forms by inward drift and the pile-up of pebbles between 3-17 au from 0.7 Myr onwards. Our results show some similarities and some key differences from these studies. We find that two reservoirs are possible if we consider a higher v$_{\rm f}$ ($\ge$ 10\,m\,s$^{-1}$), where a smaller ice-rich mass of planetesimals originating from retro-diffusion (or the cold-finger effect) can form at the snow line, and a much larger reservoir of $\sim$ 100 M$_{\oplus}$ can form within the dust trap beyond a gap. The mass of planetesimals beyond the gap is dependent on the radial location of the gap, i.e., if the gap is closer-in, planetesimal formation can start earlier and proceed for longer, resulting in more plantesimal mass formed.

Recent studies \citep{carr21,carrsimon22} argue that particles smaller than cm-sized may not lead to planetesimal formation in rings. It is therefore likely we are overestimating planetesimal formation at the dust trap, especially beyond the outermost gaps (i.e., 100 au) where particles are largely mm-sized or smaller. However, as mentioned before, planetesimal formation at the dust trap has little effect on vapor enrichment in the inner disk.

Overall, our simulations suggest that planetesimal formation is only significant if it occurs at the snow line region, and it does not matter for vapor enrichment in the inner disk if it occurs in the dust trap beyond the gap.

\subsection{Exploring multiple gaps}\label{results:numbergaps}
\begin{figure*}[htb!]
	\centering
	\includegraphics[scale=0.65]{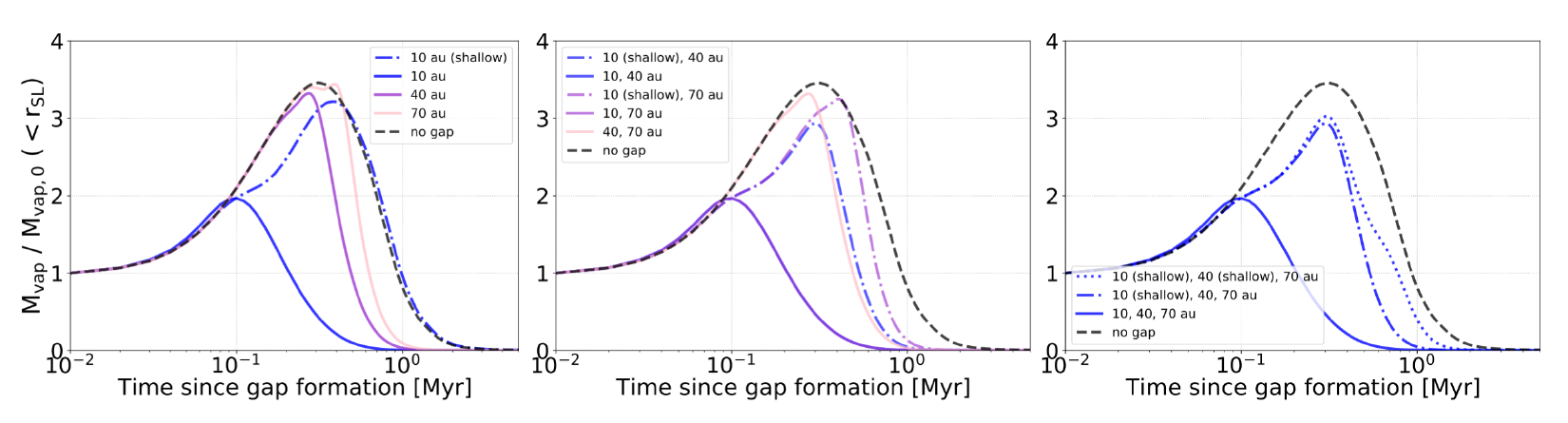}
	\caption{ Time evolution of vapor enrichment for simulations with one gap (left panel), two gaps (middle panel) and three gaps (right panel) with fiducial v$_{\rm f}$ = 5\,m\,s$^{-1}$ both inside and outside of the snow line, and with planetesimal formation. Colors denote specific gap locations; solid colored lines denote deep gaps at those locations, while dot-dashed colored lines denote shallow gap(s) at those locations. Dotted line in right panel denotes case with two shallow interior gaps, along with deep outer gap. Black dashed line shows no gap case for comparison, in each panel.}
	\label{Fig:numbergaps}
\end{figure*}

We finally also explore how the presence of additional gaps in the outer disk can affect the water enrichment in the inner disk (with planetesimal formation). From data in the observational surveys of protoplanetary disks with gaps from \citet{huang18a} and \citet{long18}, we find that gap locations span a wide radial range and peak around 40 au. We therefore select 10, 40 and 70 au as three representative radial locations to introduce one, two or three gaps in our disk simulations, as presented in left, middle and right panels of Figure \ref{Fig:numbergaps}. 

We first consider the effect of multiple gaps where all the gaps are of the same gap depth, i.e., fiducial value of 10 $\times$ $\alpha_0$. These profiles are depicted as the solid lines in the three panels in Figure \ref{Fig:numbergaps}. Focusing only on these profiles, we see that, as expected, 10 au (being the inner most gap) is the most efficient barrier relative to the other gap locations (left panel). However, it remains just as efficient a barrier as when paired with one or more outer gaps (middle and right panels). This implies that irrespective of whether additional outer gaps exist, a deep inner disk gap can be an excellent barrier to pebble delivery. In a similar vein, the 40 au gap is just as efficient a barrier whether present alone or when paired with an even outer 70 au gap. For the disk with three gaps, the additional presence of a third gap at 70 au (that would block the least amount of icy pebble material from inner disk), even if it is deep, does not matter at all.

We next consider what would happen if an inner gap is shallower in depth compared to the outer gaps in the disk. We assume a gap depth of 1.5 $\times$ $\alpha_0$ for only the innermost 10 au gap, and retain fiducial gap depth for the outer two gaps. Profiles incorporating a shallow 10 au gap are depicted with a dot-dashed profile in the same three panels of Figure \ref{Fig:numbergaps}. The left panel shows how a shallow 10 au gap is inefficient at trapping particles beyond it and continuously leaks material through the gap (as seen earlier in Section \ref{results:gapstrength}). When paired with an outer deep gap, the outer gap is able to restrict pebble delivery from beyond it, and thus reduces the total amount of material that may be leaked through the inner shallow gap. Here, as expected, the 40 au gap performs better than the 70 au gap simply because it has more material to block compared to the latter. Even here, in the case of the disk simulation with three gaps, the additional presence of a third gap at 70 au does not matter at all. Another case with two shallow inner gaps at 10 and 40 au, accompanying a deep outer gap at 70 au shows a slightly more prolonged vapor enrichment (little over 1 Myr) in the inner disk.

All these simulations suggest that water delivery to the inner disk is strongly regulated by the deepest innermost gap that is present, or in combination, the pair of deep inner most gaps that is present. 

\section{Discussion}\label{sec:discussion}

In the following section, we discuss the main insights gained from the parameter study described in detail in Section \ref{sec:results}. 

\subsection{Disk Structure and Vapor enrichment}

The overarching motivation of this work is to understand how the outer disk may affect vapor enrichment in the inner disk by altering pebble dynamics in the outer disk. Disk structure present in the form of a gap (which is the focus of this work) can block partially or completely the passage of radially drifting ice-rich material across it, which otherwise unhindered in its inward travel  would sublimate to vapor in the inner disk within the snow line. In this work, we perform several simulations where we vary the radial location of the gap, the depth of the gap (i.e., how efficiently it can filter out particles from passing through it) and even explore the influence of additional gaps on inner disk vapor enrichment. For our fiducial value of fragmentation velocity v$_{\rm f}$ = 5\,m\,s$^{-1}$ (representing some median value from all experimental results performed so far on the tensile strengths of icy and silicate particles against collisions), and for a turbulent $\alpha$ = 1 $\times$ 10$^{-3}$, we find that for a uniform disk, the inner regions can become strongly enriched in water vapor due to delivery of ice-rich pebbles from the outer disk. This enrichment can generally last for about 1 Myr or so, and has a typical profile: an increase with pebble delivery, a peak (when bulk pebble mass is delivered) and subsequent decrease with stellar accretion. For a disk with a gap, vapor enrichments may be less strong, with lower peak enrichments as compared to the initial value for a uniform disk with no gap; vapor enrichment episodes may also be more time-limited for disks with a gap, compared to disks with no gap. If all gaps are deep and are efficient barriers against pebble passage, then gaps present in the outer disk ( $\sim$ 50 - 100 au) have the disadvantage that they can only block as much material as there is beyond them. An inner gap ($\sim$ 10 au) in this way occupies a special location in the disk, simply because it is able to block most amount of ice-rich material originating from the outer disk, which would drift inwards into the inner disk. 

Gaps may not be always efficient at trapping material beyond it. The depth of the gap affects its trapping efficiency. We find that, just as before, it is the depth of the innermost gap that regulates the vapor enrichment in the inner disk. A shallow inner gap can continuously leak material from beyond it resulting in a longer and higher vapor enrichment episode in the inner disk than with a deep inner gap. Shallow outer gaps would have this effect as well, though not as strongly, as they do not block enough ice beyond them. 

For the above reasons, if there are any additional gaps that accompany an inner gap in the disk, their presence has no real effect unless the inner gaps themselves are shallow therefore making for weak traps. If that is the case, then it is the best combination of inner gaps (that together effectively trap more pebble mass) that regulates pebble delivery into the inner disk, as discussed in Section \ref{results:numbergaps}. 

Overall, we find that it is the innermost gaps that have a dominating influence in dictating pebble delivery and therefore water enrichment in the inner disk. 

\subsection{Disk Structure vs. Planetesimal Formation}
\begin{figure*}[htb!]
	\centering
	\includegraphics[scale=0.65]{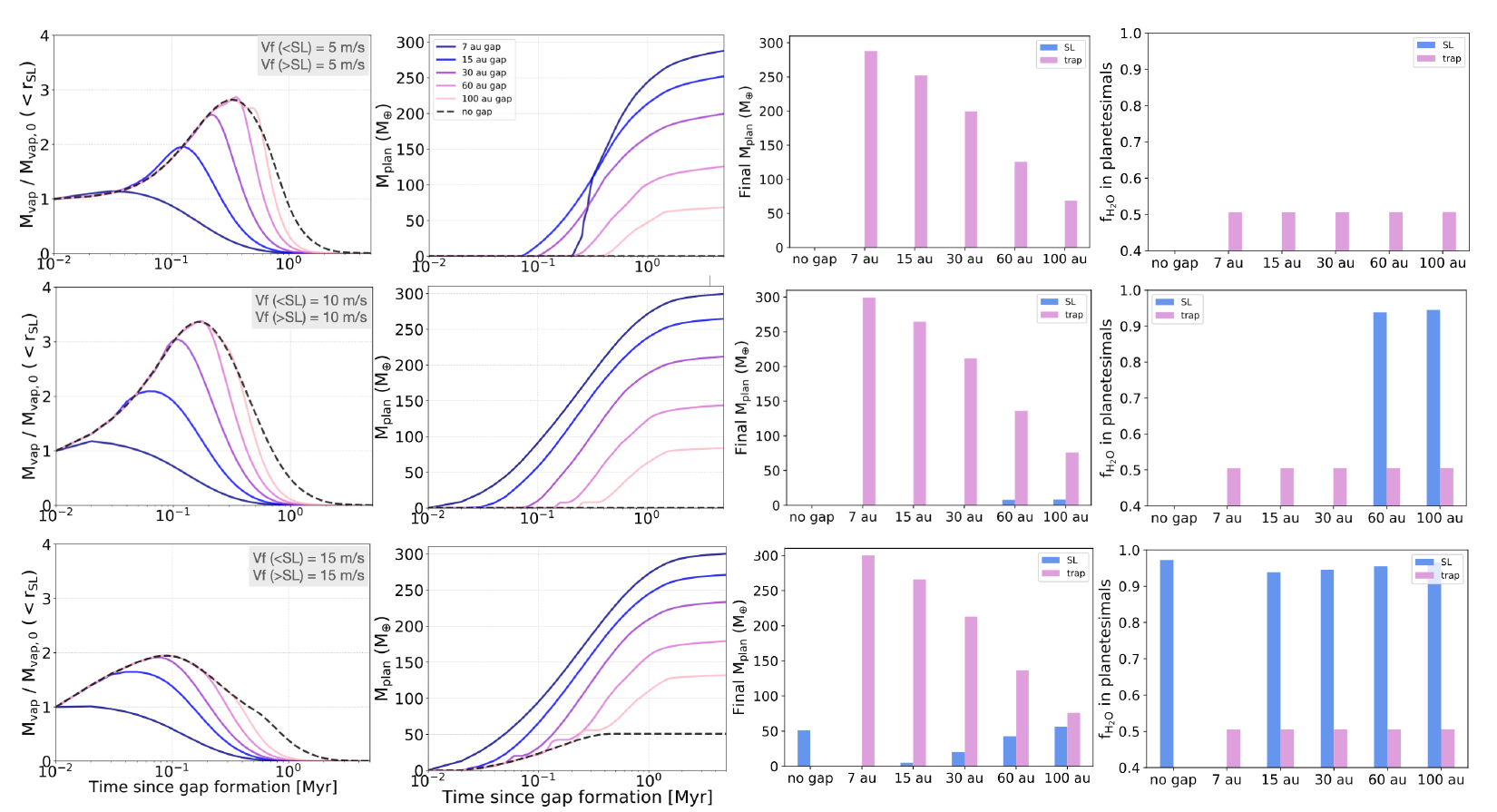}
	\caption{Plots shown are similar to Figure \ref{Fig:planform1}, but with higher initial disk mass of M$_{\rm disk}$ = 0.1 M$_{\odot}$.}
	\label{Fig:planform2}
\end{figure*}
Formation of a gap and the trapping of dust material beyond it, although extremely efficient, may not be the only way to block pebble delivery into the inner disk. As explored in other studies \citep[]{mcclure19}, it is also possible that planetesimal formation may be able to lock volatile-rich material within planetesimals and prevent them from entering the inner disk. \citet[]{najita13} also theorized that locking up of water-ice in planetesimals could be a possible reason why they found a correlation between HCN/H$_2$O ratios and dust disk mass; more massive disks may have more planetesimal formation, and therefore lock more water ice in them, leading to lower relative abundance of H$_2$O compared to HCN in the inner regions of those disks. 
In this study, we include planetesimal formation in one set of simulations to understand which of the two effects (presence of gaps or planetesimal formation) may have the more dominating impact on vapor enrichment in the inner disk. We find that for v$_{\rm f}$ = 5 or 10\,m\,s$^{-1}$, gaps have a stronger effect than planetesimal formation. When v$_{\rm f}$ $\le$ 10\,m\,s$^{-1}$, particle sizes are still limited by the fragmentation limit such that planetesimal formation mainly only takes place in the pressure bump beyond the gap. However, for v$_{\rm f}$ = 15\,m\,s$^{-1}$, particles grow to sizes large enough that a significant mass of planetesimals begin to form just beyond the snow line as well. This additional surge of planetesimal formation locks water-ice beyond the snow line and prevents its delivery into the inner disk, drastically depleting water content in the inner disk. Thus, we see that it is only for very high values of v$_{\rm f}$ ($\sim$ 15\,m\,s$^{-1}$) that planetesimal formation starts to become a dominating influence on vapor enrichment. For a more reasonable range of v$_{\rm f}$ values, pebble trapping by gap formation may still be the most efficient way to block pebble delivery into the inner disk. We perform the same simulations presented in Figure \ref{Fig:planform1} for a higher intial disk mass (M$_{\rm disk}$ = 0.1 M$_{\odot}$) shown in Figure \ref{Fig:planform2}. We find no clear trend with increasing disk mass for v$_{\rm f}$ = 5, 10 and 15\,m\,s$^{-1}$. Our simulations thus lead us to conclude that even at higher initial disk mass, pebble delivery into the inner disk is primarily affected by trapping beyond gaps rather than planetesimal formation (see also Appendix \ref{app:planform}, Figure \ref{Fig:planform_app3}).

\subsection{Fragmentation velocity vs. Turbulent viscosity}

Fragmentation velocity v$_{\rm f}$ is an important physical quantity that dictates how large particles can get before breaking apart from collisions and is determined experimentally from microgravity experiments on dust aggregates. A high v$_{\rm f}$ can enable the growth of sufficient mass of pebbles that can eventually participate in planetesimal and planet formation, while a low v$_{\rm f}$ can severely restrict the growth of particles to sub-mm sizes, which may be too small to contribute to planet formation. Recent experiments suggest a lack of consensus on the relative tensile strength of icy dust aggregates compared to bare silicates \citep{musio16,gund18,musiowurm19}
In our work, we considered both possibilities that experiments yield: either that icy silicates are more stickier than bare silicates and therefore have a higher v$_{\rm f}$ threshold, or that both are equally susceptible to collisions and have comparable v$_{\rm f}$. We therefore performed several simulations considering a range of v$_{\rm f}$. The maximum particle size at the fragmentation limit a$_{\rm frag}$ is proportional to v$_{\rm f}^2$/$\alpha$ (Equation \ref{eqn:afrag}). We therefore expanded our suite of simulations to include a range of $\alpha$ values.  

Due to the dependence on v$_{\rm f}^2$/$\alpha$, we find that our results fall into three general categories:
\begin{enumerate}

    \item \textit{low $\alpha$}: This category (top row) represents our typical simulations, showing a peak enrichment period in vapor that eventually declines with time in the inner disk at around a Myr or so for the low v$_{\rm f}$ case (1\,m\,s$^{-1}$), and around 0.5 Myr for the high v$_{\rm f}$ case ( $\geq$ 5\,m\,s$^{-1}$)). Disks with low turbulence show different extents of vapor enrichment in the inner disk depending on whether a gap is present in the outer disk, and on the radial location of the gap. 

    \item \textit{moderate $\alpha$}: This category (second and third rows) shows mixed outcomes. For $\alpha$ ranging from $\sim$ 5 $\times$ 10$^{-4}$ to 1 $\times$ 10$^{-3}$, for low v$_{\rm f}$ (1\,m\,s$^{-1}$), vapor mass slowly declines in the inner disk with time. (Note: for $\sim$ 5 $\times$ 10$^{-4}$, vapor mass stays roughly constant for a disk with no gap). On the other hand, the high v$_{\rm f}$ ( $\geq$ 5\,m\,s$^{-1}$) simulations follow the profiles in category 1 - where as discussed above, the inner disk is temporarily enriched in vapor. The extent of peak enrichment is dependent on $\alpha$ (i.e., higher $\alpha$ leads to lower peak enrichments) and whether there is a gap in the disk, and if present, where.
    
    \item \textit{high $\alpha$}: This category (bottom row) shows only a decline in mass of vapor with time over 5 Myr, with no period of enrichment like in the other cases. For higher v$_{\rm f}$ ( 10\,m\,s$^{-1}$), cases with a gap show earlier depletion of vapor within $\sim$ 1 Myr after gap formation.

\end{enumerate}

In general, if most disks are truly less turbulent than previously thought, as recent observations suggest \citep[]{pinte16,flaherty20, rosotti23}, then irrespective of v$_{\rm f}$, our simulations suggest that the inner disk will experience an intense but short-lived episode of vapor enrichment; the presence and location of the gap determining the intensity of the enrichment episode. On the other hand, disks with high turbulent $\alpha$, irrespective of v$_{\rm f}$, may not see any enrichment in vapor in the inner regions. In this way, vapor enrichment can be a diagnostic of turbulence in the disk, although a time-evolving one.

\subsection{Origin of Compact Disks}
Recent studies \citep[]{jenn22,kurt21} suggest that substructure may be common in compact disks. Compact dust disks may form from rapid dust evolution and drift from previously large disks that perhaps did not have any major substructure that hindered pebble drift \citep[][]{marelmulders21}. As discussed in detail in this study, such un-structured uniform disks likely show strong and prolonged vapor enrichment lasting $\sim$ Myr) in their inner disks. It is, however, also possible that some disks are simply born smaller in size \citep{najbergin18}. For all the simulations presented in this work, we choose the initial size of the gaseous disk to be 500 au (with a critical radius of 70 au, concentrating most disk mass within), and do not actually model small disks. However, we still attempt to predict vapor enrichment outcomes for disks that formed small, based on our study. 

The initial disk mass that would have to be assumed for modeling small disks is important. If we adopt a similar initial disk mass ($\sim$ 0.05 M$_{\odot}$) as we did for disks in this study, these small disks would be highly dense. Any gaps in such a disk would be shallower and therefore even more `leaky', and would allow for much more high and prolonged vapor enrichment in its inner regions. If we rather adopt an initial disk mass that is presumably scaled to its disk size, such a disk would be a small-scale version of the disks presented here, and would likely exhibit similar but smaller vapor enrichment profiles, as discussed in detail in this work, i.e.,  deep inner gaps in these disks can efficiently block pebble delivery (leading to low vapor enrichment) but shallow gaps may not (leading to moderate vapor enrichment).
Thus, while small uniform disks may show high and prolonged vapor enrichments if born large, we argue that even if born small, there is a possibility that small disks with inner gaps may also show high and prolonged vapor enrichments if they formed dense. 

\section{Summary and Conclusions}\label{sec:conclusions}

In this study, we aimed to understand the overall effect of disk structure on vapor enrichment in the inner disk, and determine what physical properties had most influence on the extent of enrichment. We built on our previous modeling \citep[][]{kalyaan21} and employed a multi-million year disk evolution model that incorporated the two-population model of \citet[]{birnstiel12} and included volatile transport, considering the sublimation of ice and freeze-out of vapor on solids at the snow line. Furthermore, we also included disk structure in the form of gaps that are able to trap icy pebbles and dust particles beyond them and explored in detail the effect of the presence of gaps, their radial location and depth on the vapor enrichment in the inner disk. We finally also explored the effect of planetesimal formation. We present the main highlights on our study as follows:

\begin{enumerate}
    \item The time evolution of the mass of vapor in the inner disk depends on the fragmentation velocity v$_{\rm f}$ of dust particles and turbulent viscosity $\alpha$ in the disk. 
    \item If disks are not very turbulent, i.e., $\alpha$ $\le$ 5 $\times$ 10$^{-4}$, then our simulations suggest that they likely experience a strong and prolonged episode of vapor enrichment (lasting about 1 Myr) followed by depletion of vapor from the inner disk. Furthermore,  the presence of a gap can significantly alter the extent of vapor enrichment, especially if present closer to the star ($\sim$ 7 au or 15 au).
    \item More turbulent disks, on the other hand, may only see a constant depletion in water vapor content in the inner disk, irrespective of v$_{\rm f}$; the presence of a gap or its location does not make much of a difference. 
    \item Shallow gaps may continuously leak material continuously enriching the inner disk with vapor. Ultimately, vapor enrichment is regulated by the deepest innermost gap present, or if multiple gaps are present, the pair of inner gaps that together trap most dust mass.
    \item For a reasonable range in v$_{\rm f}$ ($\le$ 10\,m\,s$^{-1}$), locking up ices in forming planetesimals beyond the snow line does not appear to have as much of an impact as the presence of gaps does in regulating vapor enrichment in the inner disk.
    \item For v$_{\rm f}$ $\ge$ 10\,m\,s$^{-1}$, planetesimal formation occurs in a few distinct locations in disks - either beyond the snow line or in dust traps beyond gaps, if gaps are present. Planetesimals formed via the cold-finger effect at the snow line are much more ice-rich (up to ice-to-rock ratios of 0.9 in our simulations) than planetesimals formed at the snow line ($\sim$ 0.5).
    \item Inner disk vapor abundance can be an important proxy for pebble mass fluxes into the terrestrial planet formation region. Although sensitive to v$_{\rm f}$ and $\alpha$, smooth disks without structure may lead to more inner disk planet formation.
\end{enumerate}

\acknowledgments
We thank the anonymous reviewer for helpful suggestions that improved the manuscript. 
A.K. and A.B. acknowledge support from NASA/Space Telescope Science Institute grant JWST-GO-01640.
Support for F.L. was provided by NASA through the NASA Hubble Fellowship grant \#HST-HF2-51512.001-A awarded by the Space Telescope Science Institute, which is operated by the Association of Universities for Research in Astronomy, Incorporated, under NASA contract NAS5-26555.
G.D.M. acknowledges support from FONDECYT project 11221206, from ANID --- Millennium Science Initiative --- ICN12\_009, and the ANID BASAL project FB210003.
M.L.\, acknowledges funding from the European Research Council (ERC Starting Grant 101041466-EXODOSS). G.R. acknowledges funding by the European Union (ERC Starting Grant DiscEvol, project number 101039651). Views and opinions expressed are however those of the author(s) only and do not necessarily reflect those of the European Union or the European Research Council Executive Agency. Neither the European Union nor the granting authority can be held responsible for them.

%

\vspace{5mm}





\appendix

\section{Gas accretion vs. Vapor Diffusion}\label{app:vapgasrates}

\begin{figure*}[htb!]
	\centering
	\includegraphics[scale=0.52]{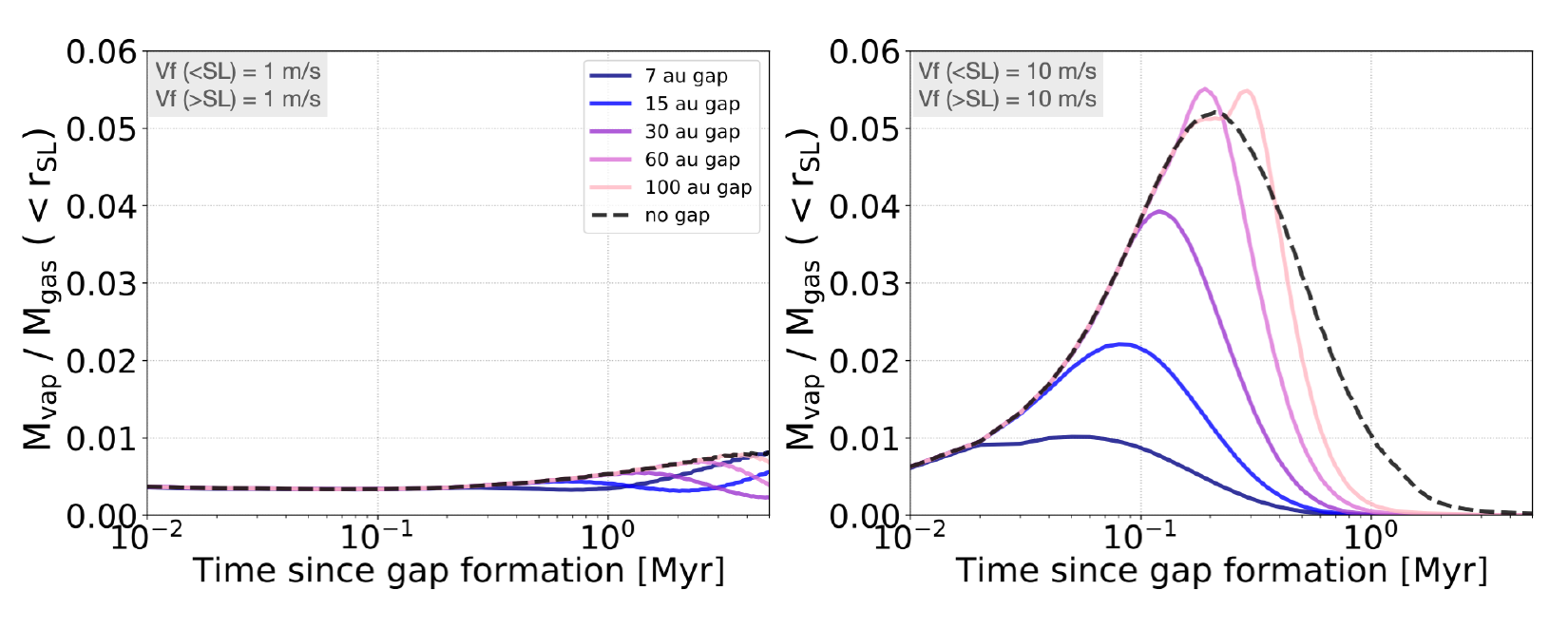}
	\caption{Plots showing `vapor abundance', i.e., ratio of  the mass of vapor within the snow line to the mass of bulk gas within the snow line, for simulations with radially uniform fragmentation velocity v$_{\rm f}$ = 1\,m\,s$^{-1}$ (left) and 10\,m\,s$^{-1}$ (right). Lines and colors are as in Figure \ref{Fig:nogrowthfrag}.}
	\label{Fig:appendix_vapgas}
\end{figure*}

\begin{figure*}[htb!]
	\centering
	\includegraphics[scale=0.25]{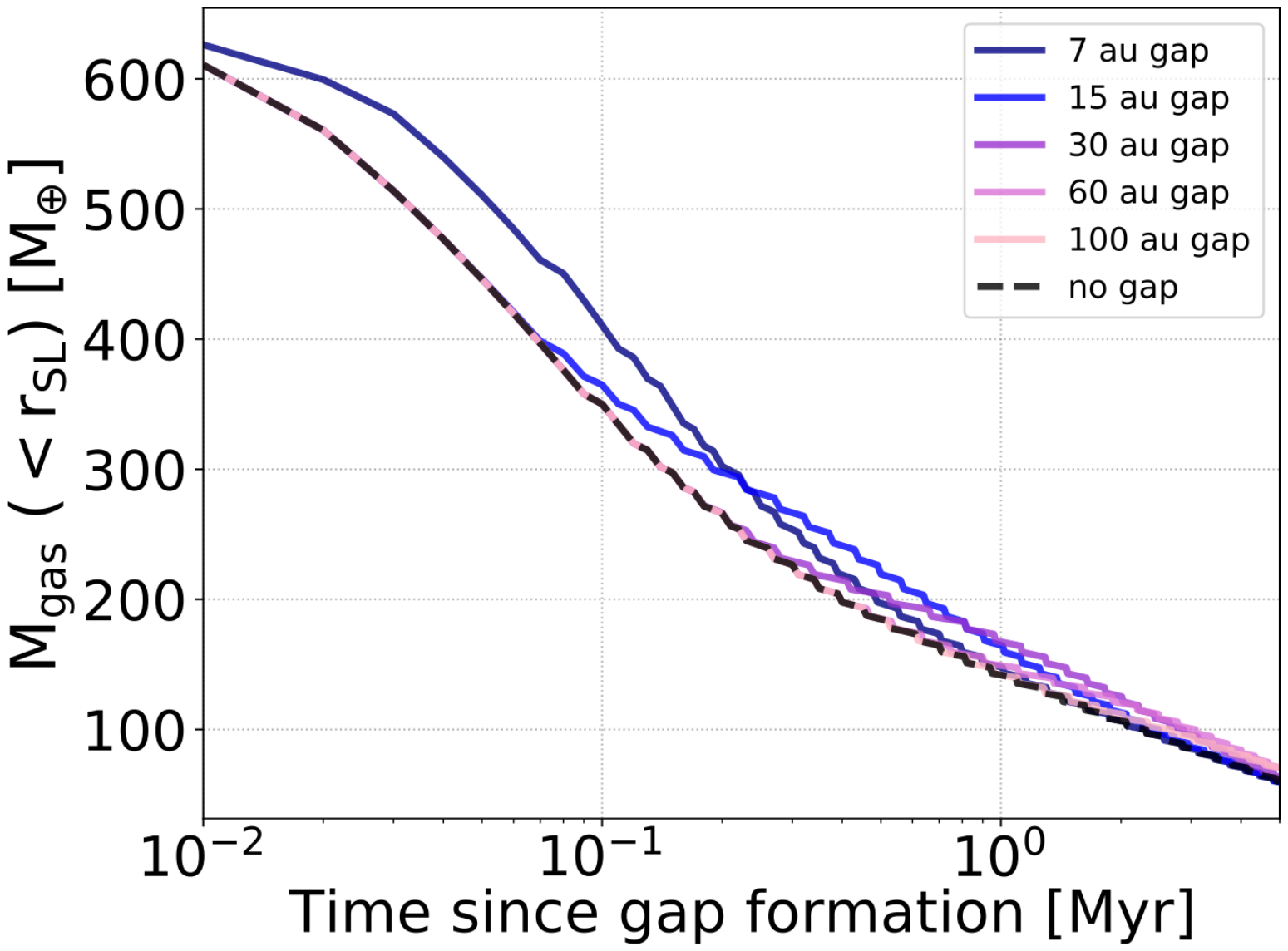}
	\caption{Plot showing the mass of gas present within the snow line over time for typical simulations with fiducial initial disk mass. Note that the snow line moves slightly inward with time.}
	\label{Fig:appendix_massgasSL}
\end{figure*}

In addition to the `vapor abundance' plots we show in Figures \ref{Fig:nogrowthfrag} and \ref{Fig:gaplocation}, we present the same plots for simulations with v$_{\rm f}$ of 1\,m\,s$^{-1}$ and 10\,m\,s$^{-1}$ (for fiducial $\alpha$) in Figures \ref{Fig:appendix_vapgas} and \ref{Fig:appendix_massgasSL} to illustrate the relative rates of gas and vapor transport in the inner disk. The mass of gas within the snow line region monotonically decreases with time as it is mainly accreted onto the star. In addition, water vapor is a trace species that advects and diffuses throughout the bulk gas. Generated at the snow line by drifting icy particles, vapor can diffuse inward and be accreted onto the star, or `retro-diffuse' again across the snow line. For 1\,m\,s$^{-1}$, our simulations show that until about 0.3 Myr, both gas and residual vapor in the inner disk are accreted onto the star at the same rate. When pebbles start to drift inward of the snow line from 0.3 Myr until 5 Myr, the inner disk gradually gets more vapor-rich, if no gap is present. If a gap is present for this case, as well as for larger v$_{\rm f}$ (5 or 10\,m\,s$^{-1}$), the inner disk is quickly flooded with vapor (from the incoming icy pebble population) which subsequently diffuses out of the region (via star or snow line) slightly more rapidly than stellar accretion.

\section{Gap depth}\label{appsec:gapstrength}
\begin{figure*}[htb!]
	\centering
	\includegraphics[scale=0.62]{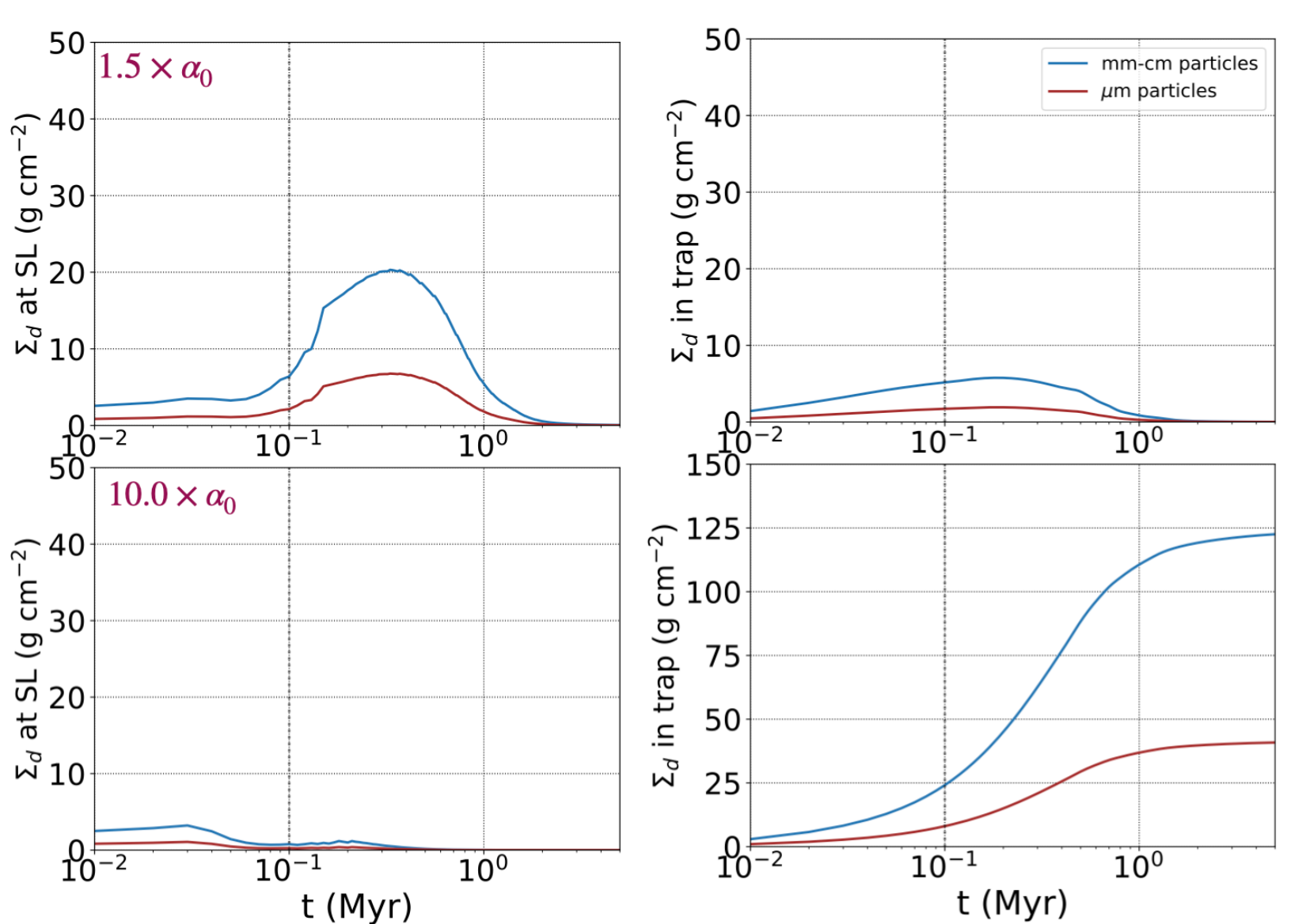}
	\caption{Dust surface densities for large (blue lines) and small particles (brown lines) at the snow line (left column) and at trap (right column) for two simulations of different gap depths for a gap at 7 au. Top row shows simulation with gap depth of 1.5 $\times$ $\alpha_0$; bottom row shows simulation with gap depth of 10 $\times$ $\alpha_0$.}
	\label{Fig:gapstrength_app1}
\end{figure*}

\begin{figure*}[htb!]
	\centering
	\includegraphics[scale=0.62]{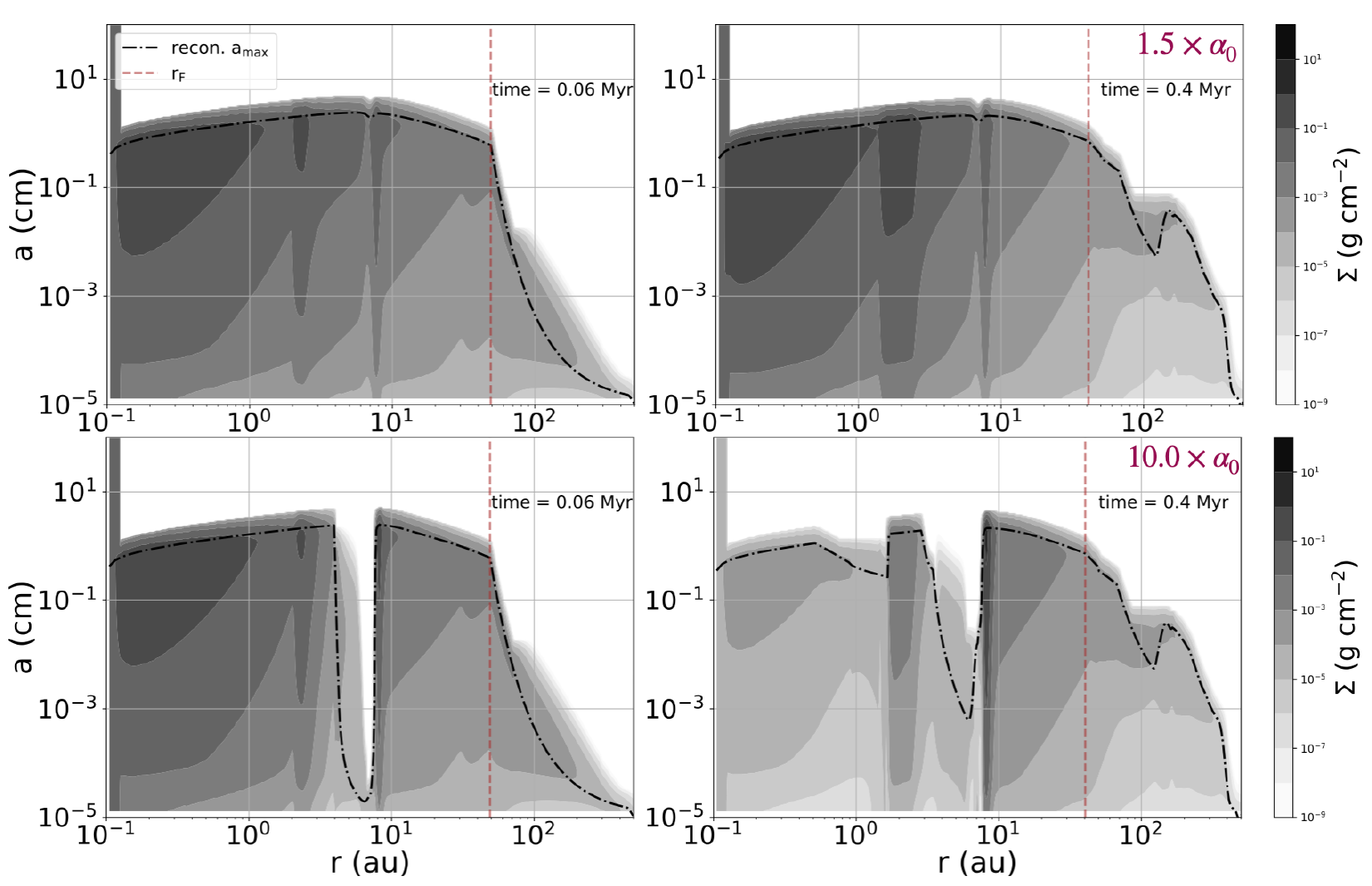}
	\caption{Dust density distributions for same simulations depicted in Figure \ref{Fig:gapstrength_app1} at two different snapshots in time: at 0.06 Myr (left column) and at 0.4 Myr (right column). Rows show simulations with shallower gap (top) or deeper gap (bottom). Black dashed line shows maximum particle size a$_{\rm max}$(r) as computed by the reconstruction model of \citet[]{birnstiel15}; vertical brown dashed line shows fragmentation radius, r$_{\rm f}$, representing the radial extent of the fragmentation dominated region in disks, also computed by the same model.}
	\label{Fig:gapstrength_app2}
\end{figure*}

In this section, we analyze two simulations performed with different gap depths in some more detail. In Figures \ref{Fig:gapstrength_app1} and \ref{Fig:gapstrength_app2}, we show two simulations of the 7 au gap with low (1.5 $\times$ $\alpha_0$) and fiducial gap depths (10 $\times$ $\alpha_0$). Figure \ref{Fig:gapstrength_app1} shows the time-evolution of the surface density of large and small particles (0.1 $\mu$m) at the snow line region and the dust trap; large and small particles here correspond to the two populations in \citet{birnstiel12} used in this work. In Figure \ref{Fig:gapstrength_app2}, we show $\Sigma$(a,r) contour plots at two specific snapshots in time (0.06 and 0.4 Myr) for the same simulations, which we generate using the reconstruction scheme developed in \citet{birnstiel15}. This work broadly demarcates the particle size - radial distance parameter space into regions where fragmentation, drift or turbulent diffusion of particles is efficient. They employ a semi-analytical treatment from which they compute a surface density for all particle size bins at each radial distance $r$ from the two- population model, effectively `reconstructing' the full dust evolution simulations from \citet{birnstiel10}. We assume that a gap being a local perturbation would not significantly affect where the global processes of fragmentation, growth and drift of dust particles across the disk are dominant. Top right panel in Figure \ref{Fig:gapstrength_app1} shows that for a low gap depth of 1.5 $\times$ $\alpha_0$, few particles are trapped and only temporarily in the pressure bump beyond the gap. Dust material is instead continuously leaking out of the trap and seen surging around 0.4 Myr at the snow line (top left panel), which eventually shows up as water vapor (Figure \ref{Fig:gapstrength}) within the snow line. Top panels in \ref{Fig:gapstrength_app2} confirm this poor trapping by showing high density of large mm-cm sized particles in the inner disk that have accumulated and grown at 0.4 Myr, compared to the very low density of such particles beyond the gap. On the contrary, for a gap depth of 10 $\times$ $\alpha_0$, large and small particles accumulate, grow and are trapped well for several Myr beyond the gap in the pressure bump (lower right panel of Figure \ref{Fig:gapstrength_app1}). The inner disk is consequently depleted at 0.4 Myr as seen in the lower right panels Figures \ref{Fig:gapstrength_app1} and \ref{Fig:gapstrength_app2}. 

\section{Leaky Gaps}\label{app:leakygaps}

\begin{figure*}[htb!]
	\centering
	\includegraphics[scale=0.62]{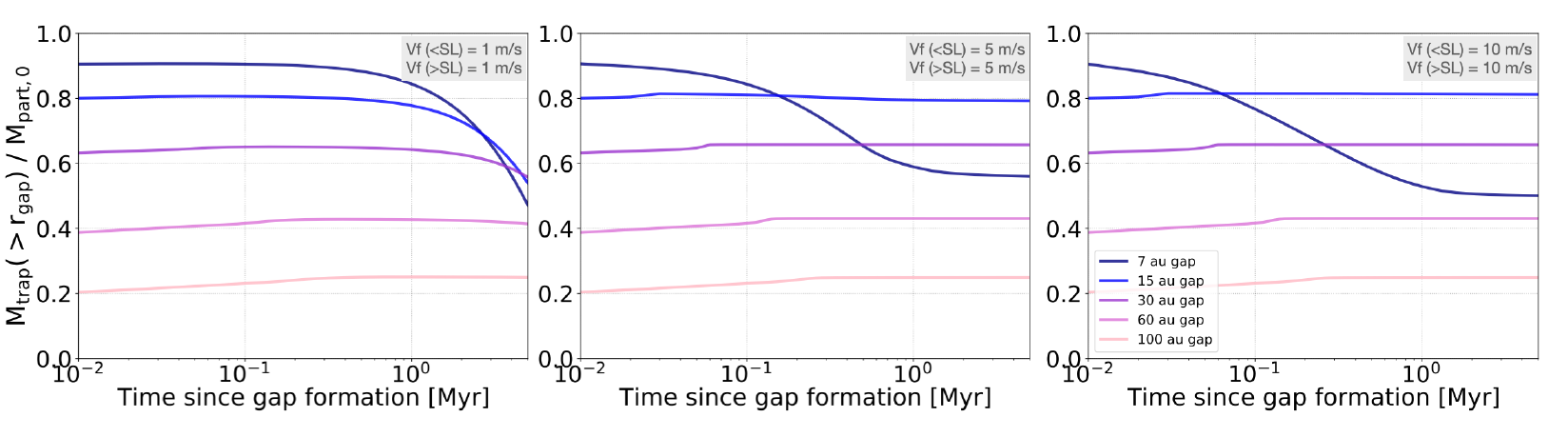}
	\caption{Plots show fraction of total initial pebble mass trapped beyond the gap at each time for simulations with v$_{\rm frag}$ = 1\,m\,s$^{-1}$ (left), 5\,m\,s$^{-1}$ (middle) and 10\,m\,s$^{-1}$ (right) respectively.}
	\label{Fig:leakygap_app}
\end{figure*}

In \citet[][]{kalyaan21} (see their Appendix B Figure 12), we show how effective different gaps are in trapping particles of different sizes beyond them. As we mentioned before in Section \ref{sec:radiallocation}, with increasing radial distance, pressure bumps are able to trap smaller and smaller particles. Therefore, for the same particle size, a similar gap would be leakier in the inner disk than in the outer disk. 

This also holds if we include growth and fragmentation of particles. We show similar plots of trapping efficiency of gaps based on their radial location in Figure \ref{Fig:leakygap_app}. For our fiducial gap depth of $\alpha$ = 10 $\times$ $\alpha_0$, we find that the innermost gaps (e.g., gap at 7 au) are slightly `leaky', as compared to outer gaps. The gap at 7 au allows the passage of small particles that are otherwise efficiently trapped beyond gaps at larger radii \citep[see also][]{stam23}.

\section{Planetesimal Formation over time}\label{app:planform}
\begin{figure*}[htb!]
	\centering
	\includegraphics[scale=0.62]{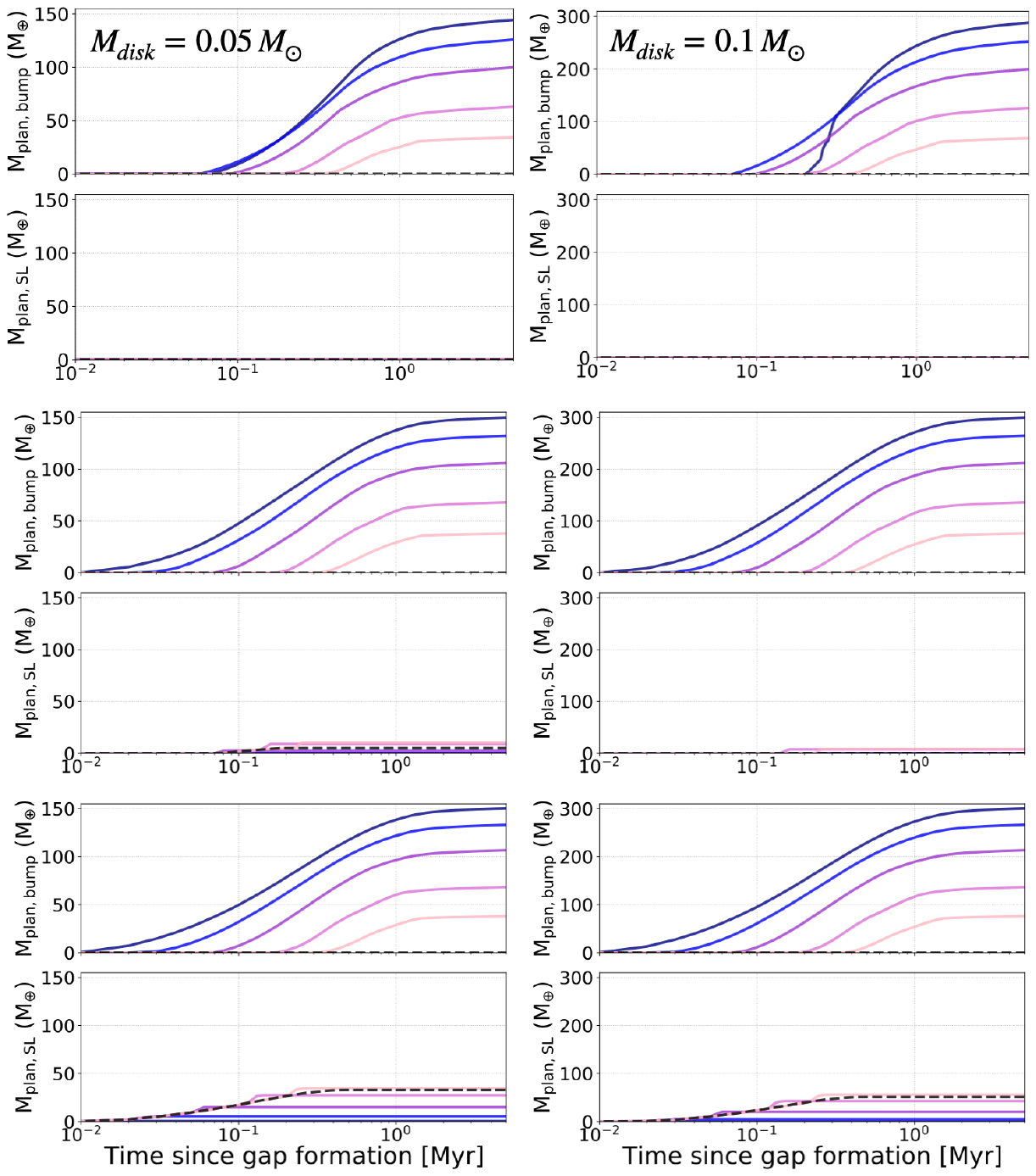}
	\caption{Figures show mass of planetesimals formed over time at either at the water snow line or dust trap (presure bump) beyond the gap, when planetesimal formation is included. Left column shows simulations with fiducial disk mass; right column shows higher disk mass. Top, middle and bottom rows represent v$_{\rm frag}$ = 5, 10 and 15\,m\,s$^{-1}$ respectively. In each case, planetesimal formation takes place for a limited duration, after which the mass of planetesimals formed remains constant with time.}
	\label{Fig:planform_app3}
\end{figure*}

In addition to Figures \ref{Fig:planform1} and \ref{Fig:planform2}, we show separately the masses of planetesimals that form either at the snow line or at the dust trap beyond the gap in Figure \ref{Fig:planform_app3}. Generally, planetesimal formation proceeds at the dust trap for a longer duration. On the other hand, the snow line region may only experience a short `burst' of planetesimal formation. Doubling the initial disk mass does not significantly affect the duration of planetesimal formation at either location.   

\bibliography{sample63}{}
\bibliographystyle{aasjournal}



\end{document}